\documentclass[12pt]{iopart}

\usepackage{iopams}
\expandafter\let\csname equation*\endcsname\relax
\expandafter\let\csname endequation*\endcsname\relax
\usepackage{amsmath,amssymb}
\DeclareMathOperator\arctanh{arctanh}
 \usepackage{graphicx}
\usepackage{color}
\usepackage{cite}
\usepackage{bm}

\def\d{\,\mathrm{d}}

\usepackage{etoolbox}
\makeatletter
\def\@mkboth#1#2{}
\newlength\appendixwidth
\preto\appendix{\addtocontents{toc}{\protect\patchl@section}}
\newcommand{\patchl@section}{%
  \settowidth{\appendixwidth}{\textbf{Appendix }}%
  \addtolength{\appendixwidth}{1.5em}%
  \patchcmd{\l@section}{1.5em}{\appendixwidth}{}{\ddt}%
}
\makeatother

\begin{document}

\title[Non-adiabatic tunneling in nanocontacts]{Quasiclassical theory of non-adiabatic tunneling in nanocontacts induced by phase-controlled ultrashort light pulses}

\author{Sangwon Kim$^{1}$, Tobias Schmude$^{2}$, Guido Burkard$^{2}$ and Andrey S. Moskalenko$^{1,2}$}
\address{$^{1}$Department of Physics, KAIST, Daejeon 34141, Republic of Korea}
\address{$^{2}$Department of Physics and Center for Applied
Photonics, University of Konstanz, D-78457 Konstanz, Germany}
\address{\mailto{$^*$moskalenko@kaist.ac.kr}}

\vspace{10pt}
\begin{indented}
\item[]\date{\today}
\end{indented}

\begin{abstract}
We theoretically investigate tunneling through free-space or dielectric nanogaps between metallic nanocontacts
driven by ultrashort ultrabroadband light pulses. For this purpose we develop a time-dependent quasiclassical theory being especially suitable to describe the tunneling process in the non-adiabatic regime, when this process can be significantly influenced by the photon absorption as the electron moves in the classically forbidden region. Firstly, the case of driving by an ideal half-cycle pulse is studied. For different distances between the contacts, we analyze the main solutions having the form of a quasiclassical wave packet of the tunneling electron and an evanescent wave of the electron density. For each of these solutions the resulting tunneling probability is determined with the exponential accuracy inherent to the method.
We identify a crossover between two tunneling regimes corresponding to both solutions in dependence on the field strength and intercontact distance  that can be observed in the corresponding behaviour of the tunneling probability. Secondly, considering realistic temporal profiles of few-femtosecond pulses, we demonstrate that the preferred direction of the electron transport through the nanogap can be controlled by changing the carrier-envelope phase of the pulse, in agreement with recent experimental findings and numerical simulations. We find analytical expressions for the tunneling probability, determining the resulting charge transfer in dependence on the pulse parameters. Further, we determine temporal shifts of the outgoing electron trajectories with respect to the peaks of the laser field in dependence on the pulse phase and illustrate when the non-adiabatical character of the tunneling process is particularly important.
\end{abstract}

%
%
%
%
%

\tableofcontents

\section{Introduction}

Together with quantum interference and entanglement, tunneling is one of the core phenomena characterizing the essence of quantum physics.
For all these three phenomena comparison to the classical description benchmarks new possibilities opening in the quantum world.
 In the case of tunneling we have also a formalism connecting both the classical and quantum description, represented by the quasiclassical Wentzel-Kramers-Brillouin (WKB) method, first proposed in a general mathematical context of linear second order ordinary differential equations \cite{Schlissel1977}. This formalism in many cases allows to obtain analytical or semi-analytical solutions and gain additional insights into their behaviour based on an extended classical intuition. The original quasiclassical approach is, however, suitable only in energy-conserving situations with static potential barriers or in some formally equivalent cases of time-dependent potentials which can be mapped to static descriptions since the time variable can be effectively seen as a spatial coordinate \cite{Dumlu2010,Dumlu2011}. The seminal work of Leonid V. Keldysh \cite{Keldysh1964_multi} established a connection between the picture of  tunneling and multiphoton ionization induced by laser fields, i.e. in temporally changing spatial potential barriers. This breakthrough achievement was followed by development of quasiclassical approaches applicable in situations of tunneling through spatial energy barriers varying with time \cite{Perelomov1966,Popov2005,LandauLifshitz3}. Thus also the regime of so-called  non-adiabatic tunneling \cite{Yudin2001,Eckle2008,Barth2011,Klaiber2015}, when a considerable energy is absorbed in the process of the underbarrier motion, could be captured within the same physical picture.
Conceptually time-dependent quasiclassical approaches can be related to the path integral formalism \cite{Feynman_book}, extended to the complex time plane and corresponding generally complex trajectories. In atomic physics, under certain conditions a formal derivation can be based on the strong-field approximation (SFA) \cite{Faisal1973,Reiss1980,Klaiber2013,Popruzhenko2014}.

Developments in ultrafast photonics leading to the appearance of tailored few-cycle laser sources opened new opportunities for studying of the control of pulse-induced dynamical tunneling from atomic systems \cite{Paulus2001,Chelkowski2004}, metal surfaces \cite{Apolonski2004} and plasmonic nanoparticles \cite{Putnam2017}. It was demonstrated that the carrier-envelope phase (CEP) of the light pulses can be utilized as a control parameter that found its practical application, e.g., as a method of the CEP measurement \cite{Paulus2003,Lemell2003,Wollenhaupt2009}. It was further shown both
experimentally and theoretically that strong ultrashort pulses can populate the conduction band in dielectrics \cite{Schultze2013} and semiconductors \cite{Schultze2014} due to non-resonant interband Landau-Zener tunneling of the electrons from the valence
band into the conduction band and generate currents in unbiased systems on ultrafast time scales \cite{Schiffrin2013,Yakovlev2015}, whereby the current direction is controlled by the CEP of the pulse.

Recently we can also observe an increased interest to tunneling in nanosystems, since it can open new perspectives for ultrafast nanoscale devices. So the quantum tunneling regime was predicted and observed for plasmonic systems with nanoscale gaps and resulting plasmonic response properties were studied \cite{Esteban2012,Savage2012,Esteban2015}. Another highly interesting case is realized in the tunneling microscope configuration where the charge transport between the tip and the surface can be strongly influenced by an external light field \cite{Bormann2010}. Moreover, application of tailored femtosecond few-cycle pulses (FCPs) in this configuration opened possibilities for sub-cycle coherent manipulation of the charge transport through the nanogap \cite{Krueger2011,Cocker2013,Yoshioka2016,Jelic2017}, enabling angstrom-scale spatial and sub-femtosecond temporal resolution in tunneling microscopy \cite{Garg2020}. In contrast to strongly spatially asymmetric configuration of the tunneling microscope,  metallic nanoantennas with nanoscale gaps represent devices having a symmetric stationary part of the spatial potential barrier so that the whole control of the charge transfer across the gap can be realized solely via the CEP-controlled FCPs or their sequences \cite{Rybka2016,Ludwig2020}.

In this work we develop a quasiclassical description of  the charge transfer between two nanocontacts induced by CEP-controlled femtosecond light pulses and determined by the probabilities of the field-driven  non-adiabatic tunneling. Therefore we consider
the tunneling process in the corresponding potentials being both space- and time-dependent and leading to values of the resulting Keldysh
parameter $\gamma$,
which determines the transition from the direct tunneling to the
multiphoton regime \cite{Keldysh1964_multi}, that can vary in a wide range. We base our theory on
a Lagrangian formulation \cite{LandauLifshitz3,Ganichev2002,Moskalenko2000,Moskalenko1999} of the quasiclassical imaginary time method (ITM)
\cite{Perelomov1966,Popov2005} that brings, in our opinion, certain advantages  with respect to the orginial ITM formulation concerning the justification of the derivation steps and finding the spatio-temporal structure of the resulting solution. With our approach we can address also the case of driving by the CEP-controlled FCPs that until now remained out of reach of the existing ITM considerations, where only solutions for even FCPs could be found \cite{Popov2001,Popov2001_JETP,Popov2004,Karnakov2009}, whereas a related approach of Keldysh provided us recently with a solution just for one case of a fixed-CEP odd single-cycle pulse \cite{Keldysh_2017}. Furthermore, the existing quasiclassical descriptions are rather suitable for free-space ionization problems whereas, as we will see, in the nanogap configuration the aspect of a small intercontact distance has to be addressed appropriately that eventually affects the solution structure.

\begin{figure}
   \centering
  \includegraphics[width=14cm]{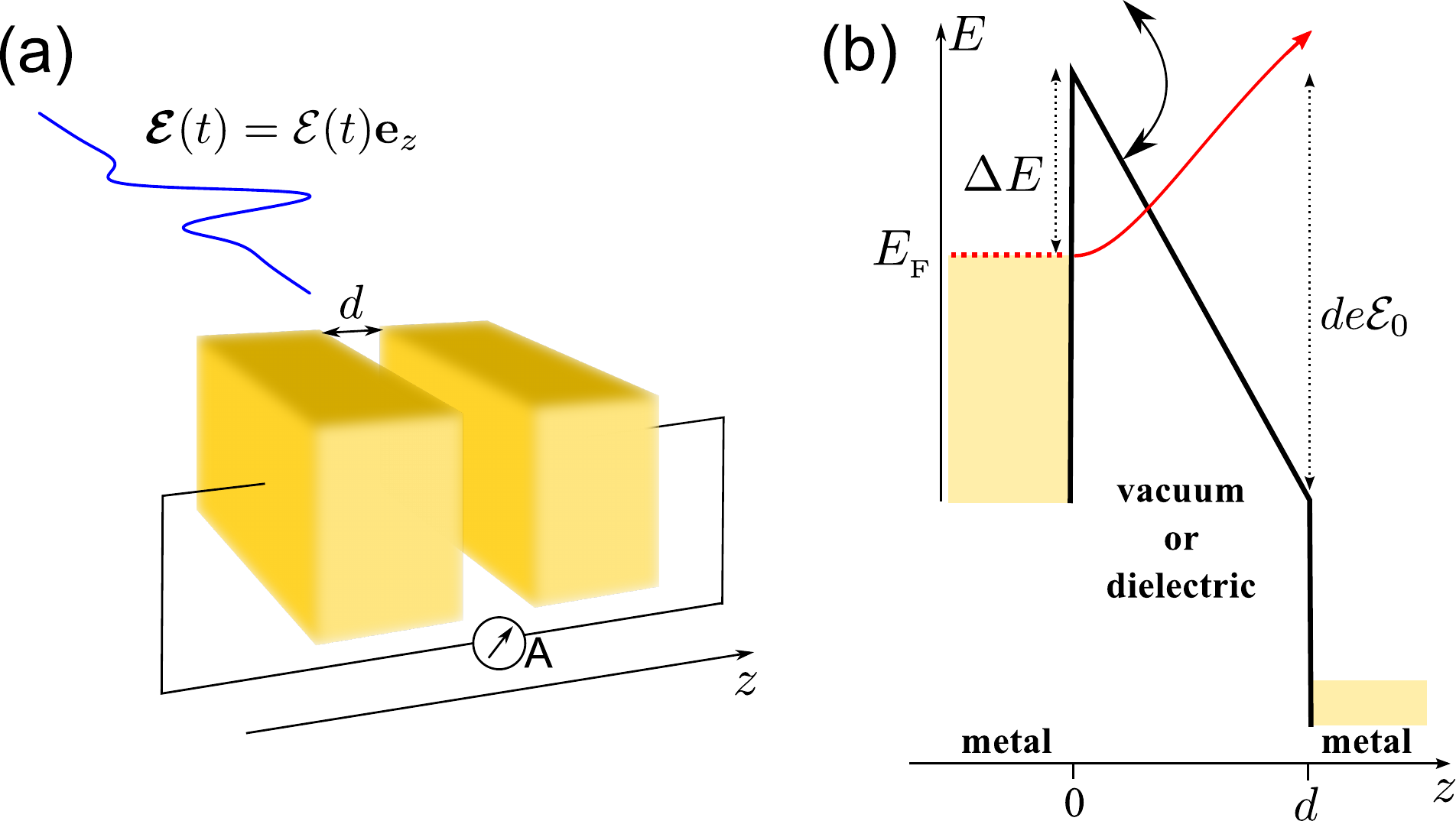}\\
  \caption{(a) Geometry of the nanogap (width $d$) formed by two metallic nanocontacts and the incoming driving few-cycle femtosecond light pulse with the electric field $\bm{\mathcal{E}}(t)$ polarized along the $z$-direction.  The nanocontacts are attached to an external electric circuit which allows to measure the electric charge transferred upon the pulse application. (b) The corresponding energy diagram for the electron travelling between the nanocontacts through the field-influenced time-dependent barrier. $\Delta E$ is the energy barrier for an electron at the Fermi level $E_{_\mathrm{F}}$ inside the metal. The red line shows an example how the energy of a tunneling electron changes with its position.
  \label{Fig:main}}
\end{figure}

\section{Quasiclassical description of non-adiabatic tunneling in ultrafast laser fields}

 \subsection{General formalism}
 We will limit our consideration to one-dimension motion of the electron along the
 coordinate $z$. In the quasiclassical approximation the electronic wave function
$\Psi(z,t)$ is given by
\begin{equation}\label{Eq:Psi_QC}
  \Psi(z,t)\propto e^{\frac{i}{\hslash}\tilde{S}(z,t)}\;,
\end{equation}
where $\tilde{S}(z,t)$ is the action. In order to find it as a
function of $z$ and $t$, one should find the general solution of
the Hamilton-Jacobi equations
\cite[footnote on p. 148]{LandauLifshitz1},
\cite[p. 298]{LandauLifshitz3},\cite{Epstein1964}
\begin{equation}\label{Eq:Hamilton_Jacobi}
  \frac{\partial \tilde{S}}{\partial t}=-\mathcal{H}(p,z,t)\;, \ \ \ \ \frac{\partial \tilde{S}}{\partial
  z}=p\;.
\end{equation}
The general solution $\tilde{S}(z,t)$ can be constructed using a
complete solution
\begin{equation}\label{Eq:HJ_complete_solution}
  S=\int_{t_0}^t
  \mathcal{L}(z',\dot{z}',t'){\rm d}t'+C=S_0+C\;,
\end{equation}
where $C$ is an arbitrary constant and
$\mathcal{L}(z',\dot{z}',t')$ is the Langrange function, as
\begin{equation}\label{Eq:tilde_S_ansatz}
  \tilde{S}=S_0+C(t_0)\;.
\end{equation}
Here $C(t_0)$ is generally an arbitrary function of $t_0$, whereas
$t_0$ has to be expressed as a function of $z$ and $t$ using the
equation
\begin{equation}\label{Eq:t0_determination}
  \left(\frac{\partial\tilde{S}}{\partial t_0}\right)_{z,t}=0\;.
\end{equation}
Taking into account $\left(\frac{\partial S_0}{\partial
t_0}\right)_{z,t}=\mathcal{H}\big|_{t=t_0}$,
equation \eqref{Eq:tilde_S_ansatz} results in
\begin{equation}\label{Eq:C_t0_determination}
  \frac{\mathrm{d}C(t_0)}{\mathrm{d}t_0}=-\mathcal{H}\big|_{t=t_0}\;.
\end{equation}
The form of the function $C(t_0)$ has to be selected in a way to
match the boundary condition on the incoming part of the wave
function at the position where the electron enters the region
under the barrier, which we assume to be located at $z=0$.
At $z=0$ and $t=t_0$, the wave function is given by
$\Psi(0,t_0)\propto
e^{\frac{i}{\hslash}\tilde{S}(0,t=t_0)}=e^{\frac{i}{\hslash}C(t_0)}$.
Comparison of this expression with the wave function in the region
outside the barrier in the neighborhood of $z=0$, up to the
exponential factors, allows to determine the form of the function
$C(t_0)$. If outside the barrier we can consider the
electron energy $E$ to be insignificantly influenced
by the perturbation of the system, then we have
\begin{equation}\label{Eq:C_t0_assumption}
  C(t_0)=-E t_0\;
\end{equation}
and equation \eqref{Eq:C_t0_determination}
turns into
$ E=\mathcal{H}\big|_{t=t_0}$\;.
This is the case for an electron tunneling between two metallic
contacts (see figure \ref{Fig:main}), because the electric
field of the pulse is screened inside the contacts, due to a quick plasmonic response. Note that it is also the
case for a charge carrier which is initially confined by a short-range
potential \cite[p.~295-297]{LandauLifshitz3},\cite{Ganichev2002,Moskalenko2000}.

The Lagrange function in equation \eqref{Eq:HJ_complete_solution}
corresponding to one-dimensional motion of the electron in the
laser electric field $\mathcal{E}(t)$ and a potential $U(z)$
%
is given by
\begin{equation}\label{Eq:Lagrange_function_def}
  \mathcal{L}(z',\dot{z}',t')=\frac{m\dot{z}'^2}{2}+z'F(t')-U(z')\;,
\end{equation}
where $F(t)=e\mathcal{E}(t)$  with $e$ denoting the electron charge and $m$ denoting its mass. $z'(t')$ satisfies the equation of motion:
\begin{equation}\label{Eq:EOM}
  m\ddot{z}'=F(t')-\frac{\mathrm{d}U(z')}{\mathrm{d}z'}\;
\end{equation}
with the boundary conditions
\begin{equation}\label{Eq:EOM_boundary_conditions}
  z'(t_0)=0\;, \ \ \ \ z'(t)=z\;.
\end{equation}
Additionally, in this case from equation \eqref{Eq:C_t0_determination}
we have
\begin{equation}\label{Eq:C_t0_determination_kinetic}
  \frac{\mathrm{d}C(t_0)}{\mathrm{d}t_0}=-\left(\frac{m v_0^2}{2}+U_0\right)\;,
\end{equation}
where $ v_0\equiv \dot{z}'(t'=t_0)$ is the initial velocity and $U_0\equiv U(z=0^{+})$ with
$z=0^+$ being the position just beyond the barrier boundary [left and right limits of $U(z)$ are different at $z=0$ in the case of an abrupt
potential step, which will be considered below]. In the case of
the tunneling from the state with energy $E$ this gives
\begin{equation}\label{Eq:t_0_determination_short_range}
  \Delta E\equiv E-U_0=mv_0^2/2\;.
\end{equation}
As far as for a tunneling state under the barrier $\Delta E<0$ [cf. figure \ref{Fig:main}(b)], we
see that $v_0$ has a purely imaginary value.

Solving the equation of motion together with the boundary
conditions allows to determine $\tilde{S}$ and
$v_0=\dot{z}'(t'=t_0)$ as functions of $z,t,$ and $t_0$. Inserting
the result for $v_0$ into
equation \eqref{Eq:t_0_determination_short_range},
we can find
then $t_0$ as a function of $z$ and $t$. In result, the action
$\tilde{S}$ can be expressed as a function of merely $z$ and $t$. Thus, using equation \eqref{Eq:Psi_QC},
we are able to find the wave function $\Psi(z,t)$. As we will see below, $t_0(z,t)$ is generally a multivalued function resulting in multiple solutions for  $\Psi(z,t)$. Within the presented formalism each of these solutions has to be analyzed separately and describes a part of the single tunneling process. However, there is no known recipe how to combine them together consistently. This issue is mostly relevant for the underbarrier dynamics close to the tunnel exit.
Otherwise,
the time-dependent probability density
at each particular point in time $t$ and space $z$
is typically dominated by just one of these solutions (see section \ref{Sec:results}). If the electric field has several oscillation cycles, there is an additional multiplication of the number of the solutions, which are separated in time approximately by the oscillation period.

Let us now assume that the electron exits the
classically forbidden region before it reaches the opposite metallic contact at $z=d$, excluding for a moment the case of very small spatial gaps between the contacts from the consideration. In order to determine the tunneling probability with exponential accuracy it is
sufficient to know the value of $|\Psi(z,t)|^2$ at the set of points in the region behind the
barrier where it has its maximum. Taking into account equation \eqref{Eq:Psi_QC}, we can see that these positions are determined by
the condition $\frac{\partial}{\partial
z}\mathrm{Im} \tilde{S}=\mathrm{Im} p=0$ and therefore
\begin{equation}\label{Eq:Im_v_exit}
  \mathrm{Im}\!\left[\dot{z}'(t)\right]=0\;.
\end{equation}
Equation \eqref{Eq:Im_v_exit} fixes the optimal classical complex trajectories \cite{Popruzhenko2014} of the tunneling electron $z_{\mathrm{opt}}(t)$. The value of $t_0$ must satisfy $z_{\mathrm{opt}}(t_0)=0$.
We see that for the optimal trajectories the electron velocity becomes real in the classically
allowed region, which is intuitively expected. For other possible trajectories the velocity generally remains complex.

From equation \eqref{Eq:Hamilton_Jacobi} follows
\begin{equation}\label{Eq:Im_v_exit2}
  \left(\frac{\partial\mathrm{Im} \tilde{S}(z,t)}{\partial t}\right)_{\mathrm{Im}p=0}=0\;,
\end{equation}
where we have a conditional partial derivative on the left hand side. This equation means that the imaginary part of the
action does not change with time along any chosen optimal trajectory when the electron moves in the classically allowed region behind the barrier. Moreover, in order to calculate the corresponding imaginary part of the action, technically we can select any arbitrary value of time $t$ on the real axis, even one corresponding to a time moment before the ultrafast laser pulse has arrived. This is possible because there is always a purely real classical trajectory  $z_\mathrm{re}(t)$ coinciding with the optimal complex trajectory for sufficiently large times after the laser pulse application, $t>t_{\mathrm{ex}}$. This real trajectory $z_\mathrm{re}(t)$ corresponds to a fictitious electron coming from the classically allowed region to the potential barrier and then reflected back. It can be obtained from $z{_\mathrm{opt}}(t)$, starting at $t>t_{\mathrm{ex}}$  and propagating back along the real time axis (cf. also \cite{Ni2016,Ni2018}).

The consideration above becomes, however, inapplicable if the width of the potential barrier becomes so small that the electron is not able to exit out of the barrier before reaching the opposite contact. For brevity, we will call this situation small-distance scenario. Equation \eqref{Eq:Im_v_exit} does not generally apply for this case since $|\Psi(z,t)|^2$ may continue to change inside the barrier with increase of $z$ up to $z=d$. To determine the tunneling probability with exponential accuracy we have to analyze the shape of $\left|\Psi(z,t)\right|^2$ at $z = d$.  Then in order to find  the optimal complex classical trajectory we need to formulate an appropriate replacement for equations \eqref{Eq:Im_v_exit} and \eqref{Eq:Im_v_exit2}.
The optimal trajectory should lead to the maximum $\left|\Psi(z,t)\right|^2$ at $z=d$ with respect to different possible values of time $t$ so that in place of equation \eqref{Eq:Im_v_exit2}  we get
\begin{equation}\label{Eq:opt_cond_case2}
  \left(\frac{\partial \mathrm{Im} \tilde{S} (z,t)}{\partial t}\right)_{z=d} = 0\;.
\end{equation}
Denoting the particular time moment when this condition is satisfied as $t_E$, in the case of the optimal complex trajectory we searching for the second of the boundary conditions \eqref{Eq:EOM_boundary_conditions} takes the form
\begin{equation}\label{Eq:EOM_boundary_conditions_case2}
  z'(t_E) = d\;.
\end{equation}



With exponential accuracy, the tunneling probability is given by
\begin{equation}\label{Eq:tun_prob_def}
  P_e=\exp(-2S_e)\;, \ \ \ S_e=\mathrm{Im} \tilde{S}/\hslash\;,
\end{equation}
where $\tilde{S}$ is evaluated at any final point $(t,z)$ belonging to the trajectory  $z_\mathrm{re}(t)$, unless the small-distance scenario is realized when we have to take $(t=t_E,z=d)$.

\subsection{Action and the propagating wave packet}\label{Sec:action_wave_packet}

Solution of the equation of motion \eqref{Eq:EOM} with the boundary conditions
\eqref{Eq:EOM_boundary_conditions} gives
\begin{align}
  z'&=v(t'-t_0)+\mathcal{Z}(t',t_0)\;, \label{Eq:EOM_solution_z}\\
  \dot{z}'&=v+\mathcal{V}(t')\;, \label{Eq:EOM_solution_zdot}
\end{align}
where the functions  $\mathcal{V}(t')$ and $\mathcal{Z}(t',t_0)$ are defined as
\begin{align}
  &\mathcal{V}(t')=\frac{1}{m}\int_{0}^{t'}F(t'') \d t''\;, \label{Eq:Dv}\\
  &\mathcal{Z}(t',t_0)=\int_{t_0}^{t'}\mathcal{V}(t'') \d t''\;. \label{Eq:Dz}
\end{align}
and the complex parameter $v\equiv\dot{z}'(t=0)$ is given by
\begin{equation}\label{Eq:v_expressed}
   v=\frac{1}{t-t_0}\left[z-\mathcal{Z}(t,t_0)\right]\equiv v(z,t,t_0)\;.
\end{equation}
Now $t_0$ in equations \eqref{Eq:EOM_solution_z}, \eqref{Eq:Dz} and \eqref{Eq:v_expressed}
is a yet unknown complex parameter that should be expressed as a function of $z$ and $t$ with help
of equation \eqref{Eq:t_0_determination_short_range}. For that a system of two real equations
\begin{align}
   &\mathrm{Re}v(z,t,t_0)+\mathrm{Re}\mathcal{V}(t_0)=0\;, \label{Eq:v_system_Re}\\
   &\mathrm{Im}v(z,t,t_0)+\mathrm{Im}\mathcal{V}(t_0)=\pm \sqrt{\frac{2}{m}\Delta E}\; \label{Eq:v_system_Im}
\end{align}
for the real $\tau_0\equiv\mathrm{Re}t_0 $ and imaginary part  $\tau_e\equiv\mathrm{Im}t_0$ of
\begin{equation}\label{Eq:t0_Re_Im}
  t_0=\tau_0+i\tau_e
\end{equation}
has to be solved.
The selection of a positive or a negative sign in equation \eqref{Eq:v_system_Im} leads
to the same physical results. Without loss of generality, we may restrict our consideration to the positive sign.
Explicit solutions can be found when the laser pulse shape $F(t)$ is fixed.

Having $t_0(z,t)$, the action $ \tilde{S}(z,t)$ can be determined using equations \eqref{Eq:HJ_complete_solution},\eqref{Eq:tilde_S_ansatz},\eqref{Eq:C_t0_assumption} and \eqref{Eq:Lagrange_function_def}.
If the the Lagrangian entering as the integrand in equation \eqref{Eq:HJ_complete_solution} does not have any singularity points as function of the complex time we may use integration by parts and write
\begin{equation}\label{Eq:S_zt_solved}
  \tilde{S}(z,t)=-\frac{m}{2}\int_{t_0}^t\left[v(z,t,t_0)+\mathcal{V}(t')\right]^2\d t'+mz\left[v(z,t,t_0)+\mathcal{V}(t)\right]-U_0t+\Delta E t_0\;.
\end{equation}
Its imaginary part is then given by
\begin{equation}\label{Eq:ImS_zt_solved}
  \mathrm{Im}\tilde{S}(z,t)=-\frac{m}{2}\mathrm{Im}\int_{t_0}^t\left[v(z,t,t_0)+\mathcal{V}(t')\right]^2\d t'+mz \mathrm{Im}v(z,t,t_0)+\Delta E \tau_e\;.
\end{equation}
Inserting  $t_0(z,t)$ into equation \eqref{Eq:S_zt_solved} and equation \eqref{Eq:ImS_zt_solved} we can determine the corresponding wave function. In particular, if it accommodates the real trajectory $z_\mathrm{re}(t)$, we obtain the shape of the propagating wave packet. The probability density is given by $|\Psi(z,t)|^2\propto e^{-\frac{2}{\hslash}\mathrm{Im}\tilde{S}(z,t)}$.

%
%
%
%

In order to discuss a more involved situation when the Lagrangian does have singularity points, let us define the  \textit{standard} path $\delta_\mathrm{s}$ in the complex time plane. This path starts at $t'=t_0$,  goes firstly parallel to the imaginary time axis from the point $(\tau_0,\tau_e)$ to $(\tau_0,0)$ and then parallel to the real time axis to $(t,0)$, thus connecting $t_0$ and $t$, whereby not crossing any singularity point. For such a path  equations \eqref{Eq:S_zt_solved} and  \eqref{Eq:ImS_zt_solved} can still be used. Moreover, these equations stay valid for any other path $\delta$ such that the closed path $\mathring{\delta}=\delta_\mathrm{s}-\delta$ does not encircle any singularity points. All these paths lead to the same result and are therefore physically equivalent.
The case of a general choice of the integration path is analyzed in \ref{App:paths} for one particular example of the pulse shape $F(t)$, which is also studied in section \ref{Sec:Ideal_HCP}. For clarity, the consideration there is limited to optimal classical complex trajectories. We found that
also for trajectories encircling singularity points  equation \eqref{Eq:ImS_zt_solved} holds. Considering the closed path $\mathring{\delta}$ mentioned above, it is helpful to introduce the winding number $n_j$ for each singularity point $j$ and the total winding number $N=\sum_j n_j$.
One can show that the standard path $\delta_\mathrm{s}$, having $N=0$, leads to the maximal possible value of the probability, whereas any paths with $N\neq 0$ give smaller values.
Physically, for a pulse-driven tunneling process the latter paths correspond to multiple reflections in the induced dynamic potential. The quasiclassical approach considered here is applicable only when the difference in probabilities between $\delta_\mathrm{s}$ and paths with $N\neq 0$ is very large (for more precise formulation, cf. \ref{App:paths}). Below we restrict our consideration to this applicability region, where it is sufficient to take into account only trajectories with $N=0$. Outside of this region the utilized quasiclassical description is not readily justified and can lead to inconsistencies   [cf. discussion after equation \eqref{Eq:P_large_gamma}].

\subsection{Optimal complex trajectory and tunneling probability}\label{Sec:conditions_optimal}
As indicated above, in order to determine the tunneling probability with exponential accuracy it is not necessary to possess information about the whole wave packet of the electron. The knowledge of $\mathrm{Im}\tilde{S}(z,t)$ along the optimal trajectory $z_{\mathrm{opt}}(t)$ (or trajectories, if there are several of them) in the region behind the barrier suffices. Therefore the task reduces to finding $z_{\mathrm{opt}}(t)$, together with the related real trajectory $z_{\mathrm{re}}(t)$, followed by calculation of $\mathrm{Im}\tilde{S}(z,t)$ for any real $t$ and $z=z_{\mathrm{re}}(t)$.

In this case the solution of the equation of motion, equations \eqref{Eq:EOM_solution_z} and \eqref{Eq:EOM_solution_zdot}, should satisfy the following conditions:
\begin{itemize}
  \item[(i)] $\mathrm{Im}[z'(t'=0)]=0,$
  \item[(ii)]  $\dot{z}'(t'=t_0)=i\sqrt{\frac{2}{m}\Delta E},$
\end{itemize}
which are also valid for any complex classical trajectory, and additionally
\begin{itemize}
  \item[(iii)] $\mathrm{Im} v=\mathrm{Im}[\dot{z}'(t'=0)]=0,$
\end{itemize}
due to  equation \eqref{Eq:Im_v_exit}. For (ii) we have selected the plus sign in front of the square root, in accordance with the convention we decided to use for equation \eqref{Eq:v_system_Im}.

The choice of $t'=0$ for (i) and (iii) is convenient but not compulsory as far as any real value of $t'$ can be taken here.
From (iii), we see directly that $v$ is real. Then,
taking into account $\Delta E>0$, these three
conditions result in
\begin{align}
  &-v\tau_e+\mathrm{Im} \mathcal{Z}(0,\tau_0+i\tau_e)=0, \label{Eq:case1conditionA}\\
  &v+\mathrm{Re}\mathcal{V}(\tau_0+i\tau_e)=0\;,
  \label{Eq:case1conditionB}\\
  &\mathrm{Im}\mathcal{V}(\tau_0+i\tau_e)=\sqrt{\frac{2}{m}\Delta E}\;, \label{Eq:case1conditionC}
\end{align}
Generally, these three equations allow us to find the
three unknown real quantities $v,\tau_0=\mathrm{Re} t_0,$ and $\tau_e=\mathrm{Im} t_0$.
When they are determined we can  express the imaginary part of the action through a real integral as
\begin{equation}\label{Eq:ImS_solved}
  \mathrm{Im}\tilde{S}=\frac{m}{2}
  \int_{0}^{\tau_e}\Big\{\left[v+\mathrm{Re}\mathcal{V}(\tau_0+i\tau)\right]^2-
  \left[\mathrm{Im}\mathcal{V}(\tau_0+i\tau)\right]^2\Big\}\d \tau+\Delta E \tau_e\;.
\end{equation}
With equation \eqref{Eq:tun_prob_def} this gives the tunneling probability. However, generally we may find multiple results for it
if there are several physically different solutions of \eqref{Eq:case1conditionA}-\eqref{Eq:case1conditionC} for $\{v,\tau_0,\tau_e\}$, e.g., as illustrated in section \ref{Sec:Realistic_Pulse}.

Let us again consider separately the small-distance scenario where the electron does not manage to exit out of the barrier into the classically allowed region before reaching the opposite contact. Proceeding as above but now taking into account equations \eqref{Eq:opt_cond_case2} and \eqref{Eq:EOM_boundary_conditions_case2}, we obtain
\begin{itemize}
  \item[(i$^\prime$)] $z'(t'=t_E) = d\;,$
  \item[(ii$^\prime$)]  $\dot{z}'(t'=t_0)=i\sqrt{\frac{2}{m}\Delta E}\;,$
  \item[(iii$^\prime$)] $\mathrm{Im} \mathcal{H}\big|_{t '= t_E}=0,$ i.e. $\mathrm{Im}\left[\dot{z}'^{\,2}(t' = t_E)\right]=0\;,$
\end{itemize}
where condition (iii) originates from equation \eqref{Eq:opt_cond_case2}. It has a clear physical meaning that at the moment the electron reaches the opposite contact it has a real value of the kinetic energy. Note that this value has actually to be negative since we have assumed that the electron does not leave the classically forbidden region. The listed conditions result finally in a system of five equations,
\begin{align}
  &(t_E - \tau_0) \mathrm{Re}v + \tau_e \mathrm{Im}v + \mathrm{Re} \mathcal{Z}(t_E,\tau_0 + i \tau_e) = d\;, \label{Eq:case2conditionA}\\
  &- \tau_e \mathrm{Re}v + (t_E - \tau_0) \mathrm{Im}v + \mathrm{Im} \mathcal{Z}(t_E,\tau_0 + i \tau_e) = 0\;, \label{Eq:case2conditionB}\\
  &\mathrm{Re} v + \mathrm{Re} \mathcal{V}(\tau_0 + i \tau_e) = 0\;, \label{Eq:case2conditionC} \\
  &\mathrm{Im} v + \mathrm{Im} \mathcal{V}(\tau_0 + i \tau_e) = \sqrt{\frac{2}{m}\Delta E}\;, \label{Eq:case2conditionD} \\
  &\mathrm{Im} v \left[\mathrm{Re} v + \mathrm{Re} \mathcal{V} (t_E)\right]=0\;, \label{Eq:case2conditionE}
\end{align}
for five real quantities: $t_E, \tau_0, \tau_e, \mathrm{Re} v, \mathrm{Im} v$. With these quantities determined, the imaginary part of the action at $z = d$ and $t = t_E$ is expressed as
\begin{align}\label{Eq:ImS_solved_case2}
   \mathrm{Im}\tilde{S}
  &=\frac{m}{2}
  \int_{0}^{\tau_e}\Big\{\left[\mathrm{Re}v+\mathrm{Re}\mathcal{V}(\tau_0+i\tau)\right]^2-
  \left[\mathrm{Im}v+\mathrm{Im}\mathcal{V}(\tau_0+i\tau)\right]^2\Big\}\d \tau \nonumber\\
  &\quad +md\mathrm{Im}v+\Delta E \tau_e\;,
\end{align}
replacing equation \eqref{Eq:ImS_solved} for the considered scenario.

\section{Results for specific pulse shapes}\label{Sec:results}
\subsection{Ideal half-cycle pulse}\label{Sec:Ideal_HCP}

In order to demonstrate in detail how the approach works for a particular
case allowing for analytical results, let us consider the following pulse shape (cf.
Refs.~\cite{Popov2001,Popov2001_JETP,Popov2005}):
\begin{equation}\label{Eq:soliton_shape}
  F(t)=F_0\frac{1}{\cosh^2\Gamma t}\;.
\end{equation}
Here the parameter $\Gamma$, having the frequency dimension,
determines the pulse duration. One can notice that the temporal integral over such a field does
not vanish, and that is generally forbidden for light pulses propagating in the far field zone. However, one might, e.g., assume that
the corresponding experimentally realized pulses possess merely weak oscillating or decaying tails \cite{Brida2014,Moskalenko2017,Arkhipov2017}. The regions of the opposite polarity in respect to the main half-cycle assure that the integral of the field over the whole time axis converges to zero.
However, the dependence of the tunneling probability on the electric field strength is highly non-linear. As a consequence, the impact
of the weak tails of the pulse on charge transfer processes governed by tunneling is negligible. Hence models like equation \eqref{Eq:soliton_shape} may be used. They can be considered for a qualitative understanding what happens during a single half cycle of a FCP.

Inserting equation \eqref{Eq:soliton_shape} into equations \eqref{Eq:Dv} and \eqref{Eq:Dz} we obtain
\begin{align}
  &\mathcal{V}(t')=\frac{F_0}{m\Gamma}\tanh\Gamma t'\;,\label{Eq:Nut}\\
  &\mathcal{Z}(t',t_0)=\frac{F_0}{m\Gamma^2}\big(\ln\cosh\Gamma t'-\ln\cosh \Gamma t_0\big).\label{Eq:Zt}
\end{align}
Here we restrict our consideration to one branch of the multi-valued complex logarithm function: its principal value (for a more general situation, see \ref{App:paths}).
Setting $t'=t$ in these equations we use them in equations \eqref{Eq:v_expressed}-\eqref{Eq:v_system_Im}.
This leads to an equation for complex $t_0=\tau_0+i\tau_e$ (or a system of two equations
for  real $\tau_0$ and $\tau_e$):
\begin{equation}\label{Eq:t0_z_t_sech}
   \frac{1}{\Gamma t-\Gamma t_0}\left[\frac{z}{z_0}-\ln\cosh\Gamma t+\ln\cosh \Gamma t_0\right]+\tanh\Gamma t_0 =i\gamma_\mathrm{_{HCP}}\;,
\end{equation}
providing us with $t_0(z,t)$. Here
\begin{equation}\label{Eq:z0}
   z_0=\frac{F_0}{m\Gamma^2}
\end{equation}
is a characteristic length and
\begin{equation}\label{Eq:gamma_Keldysh_HCP}
   \gamma_\mathrm{_{HCP}}=\frac{\Gamma}{F_0}\sqrt{2m\Delta E}
\end{equation}
is a generalized Keldysh parameter \cite{Keldysh1964_multi,Popov2004,Popruzhenko2014,Landsman2015} for the case of ideal half-cycle pulses.

\begin{figure}
   \centering
  \includegraphics[width=\textwidth]{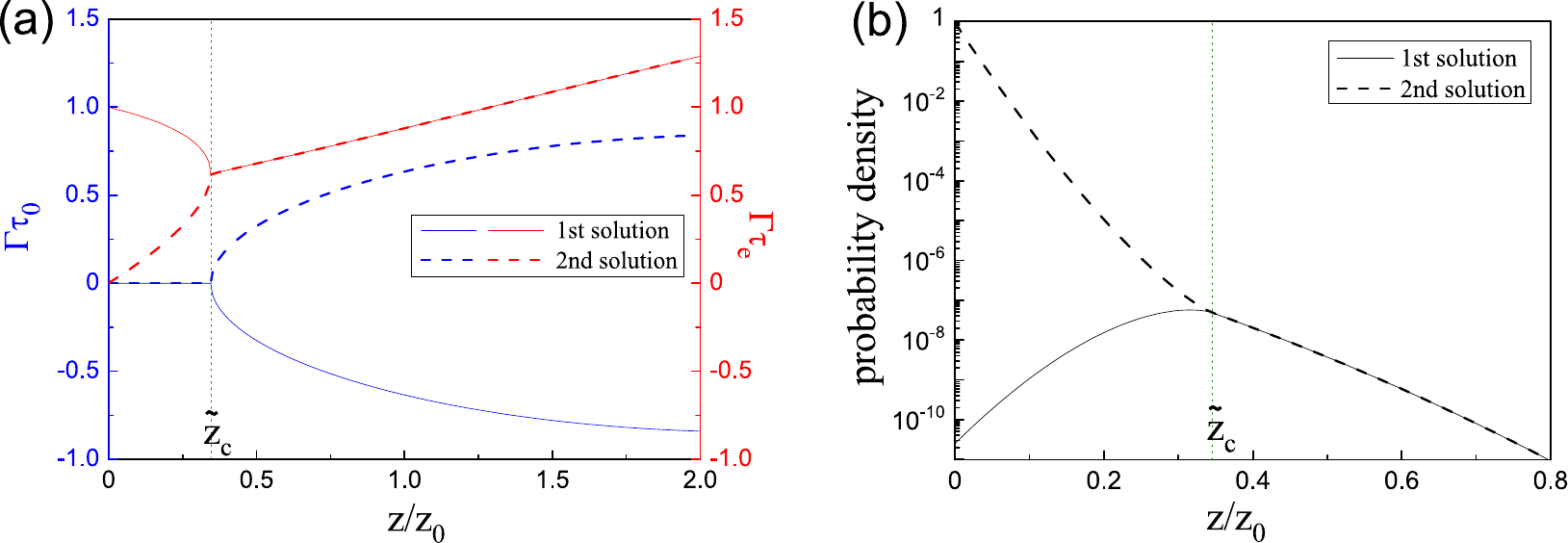}\\
  \caption{(a) Normalized values of $\tau_0(z,t=0)$ and $\tau_e(z,t=0)$ corresponding to the two main solution branches in dependence on the position $z$. At the selected time moment $t=0$ the electric field of the applied pulse reaches its maximum. The solutions are found from  equation \eqref{Eq:t0_z_t_sech} for $\gamma_\mathrm{_{HCP}}=0.94$. $\tilde{z}_\mathrm{c}\equiv z_\mathrm{c}/z_0$, where $z_0$ is given by equation \eqref{Eq:z0}. (b) Corresponding tunneling probability obtained using equation \eqref{Eq:Im_S_HCP_WP} is shown in dependence on the position $z$, with $F_0^2/(\hbar m \Gamma^3)=34.7$. }
  \label{Fig:bifurcation_plot}
\end{figure}

\begin{figure}[t!]
   \centering
  \includegraphics[width=\textwidth]{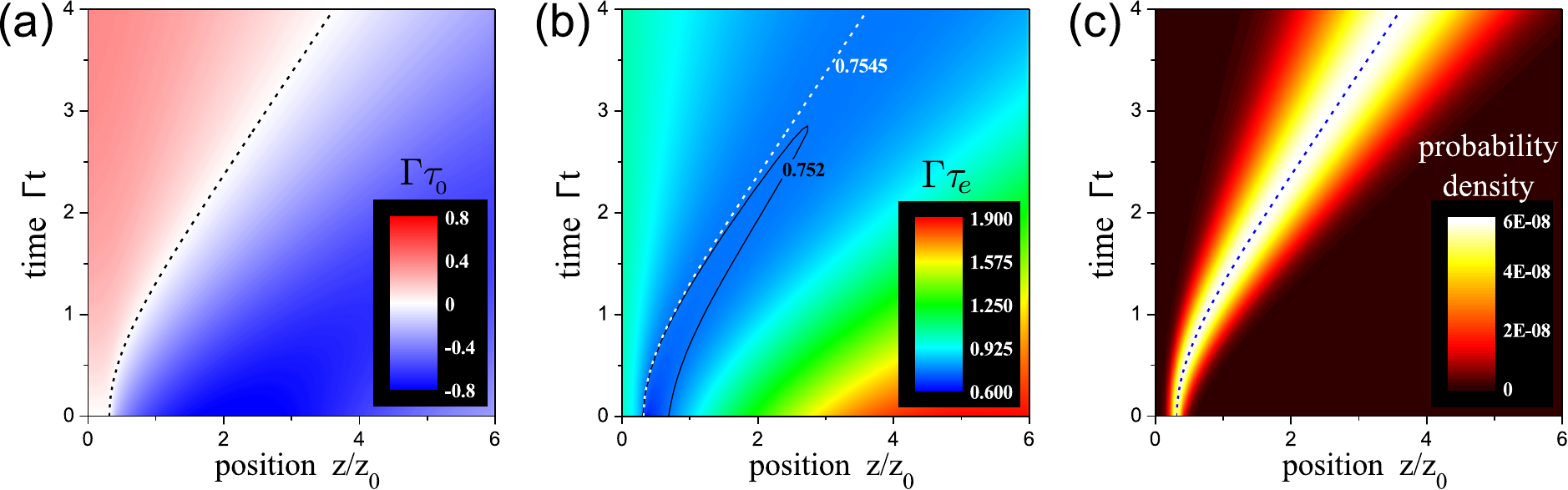}\\
  \caption{First solution branch. Starting time of tunneling $\tau_0=\mathrm{Re}t_0$ (a) and imaginary tunneling time $\tau_e=\mathrm{Im}t_0$ (b)  are multiplied by $\Gamma$ (for normalization) and  shown as functions of final real coordinate $z$ and time $t$. They were determined from equation \eqref{Eq:t0_z_t_sech} for $\gamma_\mathrm{_{HCP}}=0.94$.  (c) Resulting time-dependent distribution of the tunneling probability $e^{-\frac{2}{\hslash}\mathrm{Im}\tilde{S}(z,t)}$ evaluated using equation \eqref{Eq:Im_S_HCP_WP} with $F_0^2/(\hbar m \Gamma^3)=34.7$.
  Dashed lines in all plots indicate the real classical trajectory $z_\mathrm{re}(t)$ of a fictitious reflected electron coinciding with the optimal complex trajectory $z_\mathrm{opt}(t)$ for real values of $t$. Along this trajectory $\tau_0=0$, $\tau_e\approx 0.7545$ are determined by equation \eqref{Eq:tau_e_result_case_1} and $e^{-\frac{2}{\hslash}\mathrm{Im}\tilde{S}}\equiv P_e \approx 5.7\times 10^{-8}$ is given by equation \eqref{Eq:P_case1}. Black contour line in (b) corresponds to $\tau_e=0.752<0.7545$ and encircles a region with even lower values of $\tau_e$ than for the optimal complex trajectory.}
  \label{Fig:tau_WP}
\end{figure}

\begin{figure}[t!]
   \centering
  \includegraphics[width=\textwidth]{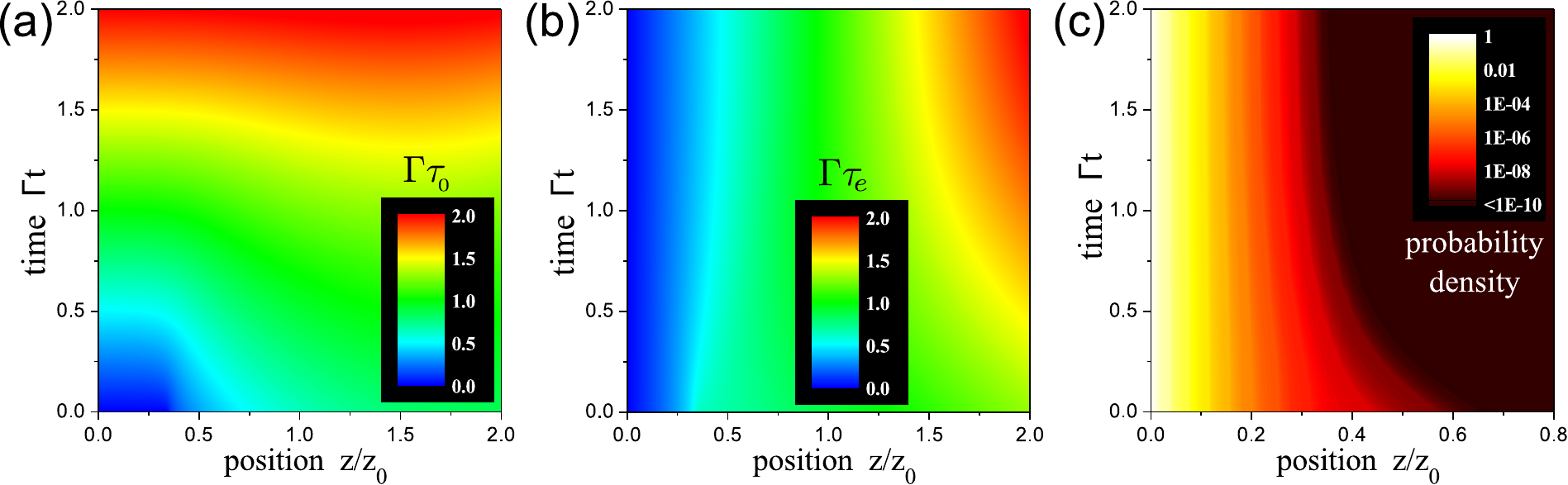}\\
  \caption{Second solution branch for the same multi-valued functions as in figure~\ref{Fig:tau_WP} and the same calculation parameters. There is no 
  outgoing real classical trajectory
  for this solution.}
  \label{Fig:tau_WP2}
\end{figure}

 As mentioned in the previous section, we get several solutions for $t_0$ from equation \eqref{Eq:t0_z_t_sech} for each pair of values of $z$ and $t$ (even though we have limited the consideration to the principal value of the complex logarithm function). The behaviour of these solutions in dependence on the position in the $(z,t)$ plane is analyzed in \ref{App:roots}. ´For the considered type of the driving field there are two physical solutions which give dominant contributions to the resulting probability distribution. Other solutions may be neglected. There is a branch point of order 1 located at $(z=z_\mathrm{c},t=0)$. By going around this point in the plane $(z,t)$ one of the two main solutions for $t_0$ transforms into another. Behaviour of  the multi-valued functions $\tau_0(z,t)$ and $\tau_e(z,t)$, limited to the two main solutions, along the line $t=0$ is illustrated in figure \ref{Fig:bifurcation_plot}(a). Their topological structure close to the point $(z=z_\mathrm{c},t=0)$ is similar to the Riemann surfaces for the real and imaginary parts of complex function $\sqrt{\zeta}$ in the neighborhood of the branch point $\zeta=0$.
 We can separate the two main solution branches from each other in order to obtain distinct single-valued functions.
 Dependencies $\tau_0(z,t)$ and $\tau_e(z,t)$ are illustrated in figures \ref{Fig:tau_WP}(a) and \ref{Fig:tau_WP}(b) for one of the main solution branches (let us name it \textit{first solution} here) and the region $t>0$. The position and time are normalized by  $z_0$ and $\Gamma$, respectively. The  dependencies $\tau_0(z,t)$ and $\tau_e(z,t)$ for the other main solution branch (\textit{second solution}) are shown in figures \ref{Fig:tau_WP2}(a) and \ref{Fig:tau_WP2}(b). As we will discuss below, the second solution dominates for the spatial region where the electron is under the barrier as well as in a certain vicinity after it may exit from this region. This solution, however, does not possess the corresponding real classical trajectory $z_\mathrm{re}(t)$ and decays at larger distances at all times $t$ , for which then the first solution overtakes the leading role.

Using equations \eqref{Eq:v_expressed},\eqref{Eq:Nut} and \eqref{Eq:Zt} in equation \eqref{Eq:ImS_zt_solved}, we obtain
\begin{equation}\label{Eq:Im_S_HCP_WP}
   \frac{2}{\hslash}\mathrm{Im}\tilde{S}(z,t)=\frac{F_0^2}{\hslash m \Gamma^3}\xi(z,t),
\end{equation}
where the dimensionless function $\xi(z,t)$ is given by
\begin{equation}\label{Eq:Im_xi_HCP_WP}
  \begin{split}
    \xi(z,t)=&\gamma^2 \Gamma\tau_e+\Gamma\!\int_0^{\tau_e}\!\!\mathrm{d}\tau\left[s_1^2(\tau_0,\tau)-
    s_2^2(\tau_0,\tau)\right]\\
    &+\mathrm{Im}\left\{\frac{\left[z/z_0-\ln\cosh(\Gamma t)+\ln\cosh(\Gamma t_0)\right]^2}{\Gamma t- \Gamma t_0}\right\}.
  \end{split}
\end{equation}
Here
\begin{align}
     &s_1(\tau_0,\tau)=\frac{1}{2}\frac{\sinh(2\Gamma\tau_0)}{f(\Gamma\tau_0,\Gamma\tau)}, \label{Eq:s1_HCP}\\
   &s_2(\tau_0,\tau)=\frac{1}{2}\frac{\sin(2\Gamma\tau)}{f(\Gamma\tau_0,\Gamma\tau)}, \label{Eq:s2_HCP}
\end{align}
with
\begin{equation}\label{Eq:f_HCP}
    f(x,y)=\cosh^2\!x\cos^2\!y+\sinh^2\!x\sin^2\!y.
\end{equation}
Everywhere $t_0\equiv t_0(z,t)$ [$\tau_e\equiv \tau_e(z,t)$, $\tau_0\equiv \tau_0(z,t)$], as it is determined by equation \eqref{Eq:t0_z_t_sech}. The dynamics of the probability distribution following from equation \eqref{Eq:Im_S_HCP_WP} is illustrated in figure \ref{Fig:tau_WP}(c) for the first solution and in figure \ref{Fig:tau_WP2}(c) for the second solution. Both probabilities are plotted together at $t=0$ in figure \ref{Fig:bifurcation_plot}(b), where we can see that they actually coincide for sufficiently large $z$ if this particular time moment is considered.
Concerning the first solution, notice that the maximum probability for any fixed time moment $t$ remains the same. It is also preserved along the optimal trajectory for real $z$ and $t$.
Thus if we view the temporal evolution of the spatial probability distribution shown in  figure \ref{Fig:tau_WP}(c)  as
the dynamics of the wave packet of the emitted electron we should take into account that this wave packet is not normalized in a way that the particle number is conserved.
Clearly, such a normalization is beyond the exponential accuracy of the applied method.
It may be partly recovered going beyond the zeroth order of
the saddle-point approximation used in the derivation of the standard imaginary time method based on the strong-field approximation \cite{Klaiber2013,Popruzhenko2014}.
Alternatively, the wave packet can be normalized to match the total probability resulting from its form at a fixed time moment, e.g. at $t=0$.
Looking at figures \ref{Fig:tau_WP}(a)-(c) we can see that along the real classical trajectory we have $\tau_0(z,t)\equiv 0$, $\tau_e(z,t)\equiv \mathrm{const}$ and $\mathrm{Im}\tilde{S}(z,t)\equiv \mathrm{const}$. $\mathrm{Im}\tilde{S}(z,t)$ reaches its maximum value on this trajectory. $\tau_e(z,t)$ has there a conditional minimum value, under the condition that this value stays possible for any $t$, including $t\rightarrow \infty$. In fact, on the real classical  trajectory $\tau_e=\displaystyle{\min_z}\; \tau_e(z,t)\big|_{t\rightarrow \infty}$. Smaller values of $\tau_e(z,t)$ are possible for finite $t$ --- see the region encircled by the solid black line in figure \ref{Fig:tau_WP}(b), but they disappear as $t$ grows.
In contrast to the first solution, the probability distribution corresponding to the second solution has an evanescent character away from the barrier, as can be observed in figure \ref{Fig:tau_WP2}(c). The values of the probability at each time moment increase along with the electric field of the applied pulse so that the probability density is instantaneously dragged in the direction of the opposite contact.

To find the corresponding optimal complex trajectory, assuming that the electron exits into the classically allowed region before reaching the opposite contact, we use equations \eqref{Eq:Nut} and \eqref{Eq:Zt} for $t'=0$ and write the real and imaginary parts  at this time moment explicitly:
\begin{align}
  &\mathcal{V}(\tau_0+i\tau_e)=z_0\Gamma
  \frac{\sinh(2\Gamma\tau_0)+i\sin(2\Gamma\tau_e)}
       {2f(\Gamma\tau_0,\Gamma\tau_e)}\;,\label{Eq:Nu}\\
  &\mathcal{Z}(0,\tau_0+i\tau_e)=-z_0\left[\frac{1}{2}\ln f(\Gamma\tau_0,\Gamma\tau_e)
  +i\arg\left(\cos\Gamma\tau_e\cosh\Gamma\tau_0+i\sin\Gamma\tau_e\sinh\Gamma\tau_0\right)\right].
  \label{Eq:Z}
\end{align}
Let  us find solutions with the absolute value of $\Gamma\tau_e$ being below $\pi/2$ (for other possible solutions as well for a clarification when and why they can be neglected, see \ref{App:paths}).
Inserting  equations \eqref{Eq:Nu} and
\eqref{Eq:Z} into equations \eqref{Eq:case1conditionA}-\eqref{Eq:case1conditionC} and eliminating the variable $v$ leads to
two equations for the determination of two real
quantities $\tau_0$ and $\tau_e$:
\begin{align}
  &\Gamma\tau_e\sinh(2\Gamma\tau_0)-2f(\Gamma\tau_0,\Gamma\tau_e)\arctan(\tan\Gamma\tau_e\:\tanh\Gamma\tau_0)=0\;,
  \label{Eq:first_tau_0_tau_e}\\
  &\frac{\sin(2\Gamma\tau_e)}{2f(\Gamma\tau_0,\Gamma\tau_e)}=\gamma_\mathrm{_{HCP}}\;.
  \label{Eq:second_tau_tau_e}
\end{align}
Generally, such a problem should be treated numerically. Here the solution is simplified by the fact
that the left hand side of equation \eqref{Eq:first_tau_0_tau_e} vanishes only if either $\tau_0= 0$ or $\tau_e=
0$. For $\tau_e=0$ equation \eqref{Eq:second_tau_tau_e} cannot be
satisfied. Thus we must have $\tau_0=0$ and therefore
$f(\Gamma\tau_0,\Gamma\tau_e)=\cos^2\Gamma\tau_e$. Then from
equations \eqref{Eq:case1conditionB} and \eqref{Eq:Nu} follows $v=0$,
i.e. the electron leaves the classically forbidden region with
zero velocity. Notice that this property is not assumed but does follow from the derivation.
As a result equation \eqref{Eq:second_tau_tau_e} simplifies to
\begin{equation}\label{Eq:tau_e_result_case_1}
  \tan(\Gamma\tau_e)=\gamma_\mathrm{_{HCP}} \;.
\end{equation}
For fixed values of $\gamma_\mathrm{_{HCP}}$ and $\Gamma$ this equation has only
one solution in the range $\Gamma\tau_e\in(-\pi/2,\pi/2)$ giving
$\tau_e=\frac{1}{\Gamma}(\arctan\gamma_\mathrm{_{HCP}})$.

\begin{figure}
   \centering
  \includegraphics[width=15.5cm]{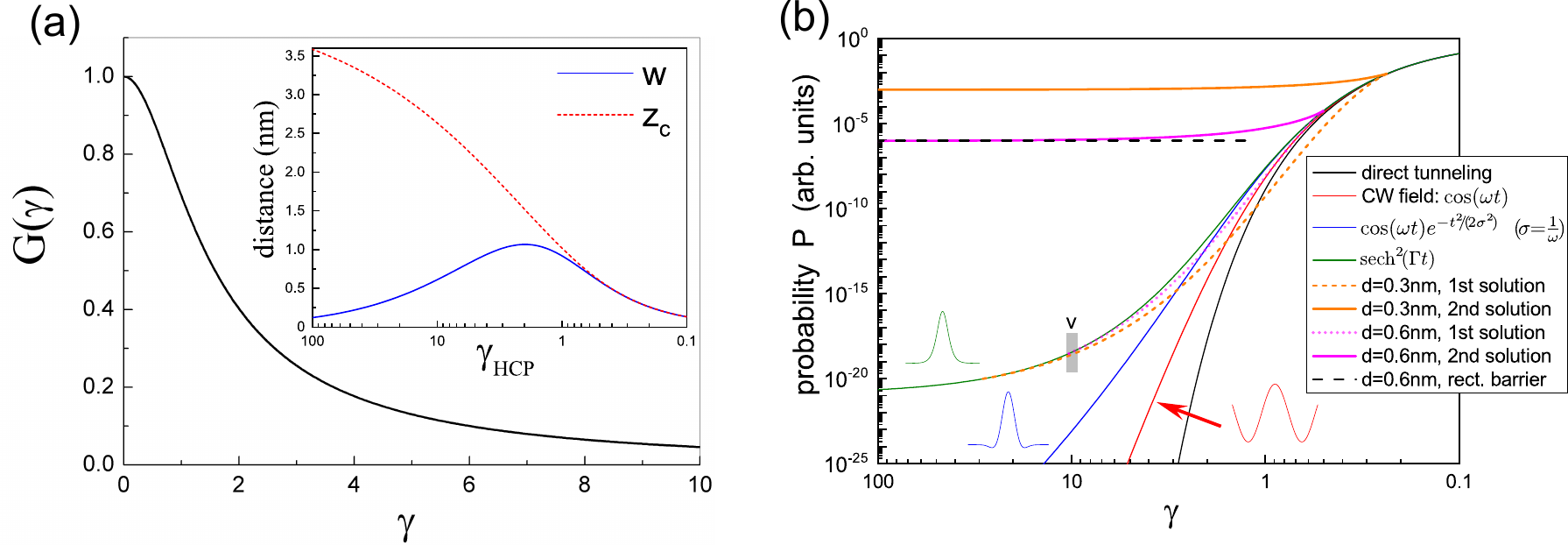}\\
  \caption{(a) Function $G(\gamma)$ entering condition \eqref{Eq:d_crossover}. Inset shows the dependence of the distance travelled by the electron under the barrier $w$ given by equation \eqref{Eq:tunneling_distance_case1} and critical distance $z_\mathrm{c}$ when $\gamma_{_\mathrm{HCP}}$ is varied in the same way and for the same conditions as described below.
  (b) Dependence of the tunneling probability on the Keldysh parameter. Green line shows the case of an ideal half-cycle pulse, equation \eqref{Eq:P_case1}.  We start from the parameters as in figures \ref{Fig:bifurcation_plot} and \ref{Fig:tau_WP}, assume $\gamma=\gamma_{_\mathrm{HCP}}$ and then vary $\gamma$ by altering the value of the peak electric field strength. Gray vertical bar with label ''v'' indicates the validity threshold \eqref{Eq:P_large_gamma_validity_threshold}. To the left of this bar we cannot rely on the result represented by the green line.
  Blue line illustrates the case of a few-cycle pulse with a Gaussian envelope with the temporal shape given by $\cos(\omega t)e^{-t^2/(2\sigma^2)}$, whereas we selected $\sigma=1/\omega=\Gamma$, connecting the time scales to the case of an ideal half-cycle pulse. Red line shows the case of continuous wave (CW) driving, also with $1/\omega=\Gamma$. In both latter cases, the calculation was based on the results given in \cite{Popov2004}. Full black line corresponds to the approximation given by equation \eqref{Eq:P_case1_small_gamma} (direct tunneling limit).  The remaining color lines show the results corresponding to the first and second solutions for $d=0.3$~nm and $d=0.6$~nm. To have a better overview, the first solution is plotted only when it differs from the case $d=\infty$ (green line) whereas the second solution is plotted only when it exceeds the first solution. For comparison, for $d=0.6$~nm also the quasiclassical probability in the case of a static rectangular barrier of this width and height $\Delta E=5.5$~eV is indicated by the dashed black horizontal line.
  }
  \label{Fig:figure_validity}
\end{figure}

Having obtained $\tau_0,\tau_e$ and $v$, we can
then determine the full optimal complex trajectory:
\begin{align}
  &z_\mathrm{opt}(t)=\mathcal{Z}(t,i\tau_e)=z_0\left[\ln\cosh(\Gamma
  t)-\ln\cos(\Gamma\tau_e) \right]
  \;.\label{Eq:trajectory_z_case_1}\\
  &\dot{z}_\mathrm{opt}(t)=\mathcal{V}(t)= z_0 \Gamma\tanh(\Gamma t)
  \;,\label{Eq:trajectory_zdot_case_1}
\end{align}
In case of the standard path $\delta_\mathrm{s}$ in the complex time plane, we can
divide the trajectory into two parts: (1) underbarrier motion $t$ where changes from $i\tau_e$
to 0 and (2) the following motion in the classically allowed region where  $t$ is real and
$t>\tau_0=0$.
For part (2) we find that $z_\mathrm{opt}>w$ holds,
where
\begin{equation}\label{Eq:tunneling_distance_case1}
  w= z_0 \ln\left[\frac{1}{\cos(\arctan\gamma_\mathrm{_{HCP}})}\right]\;
\end{equation}
is the distance travelled by the electron under the barrier. Considering equation \eqref{Eq:trajectory_z_case_1}
only for real $t$ but extending it to negative values, we get the real trajectory $z_\mathrm{re}(t)$ of the fictitious electron reflected from the barrier
(see figure \ref{Fig:tau_WP}). It is interesting to notice that the value of $w$, somewhat counterintuitively, does not coincide with the distance to the branch point $z_\mathrm{c}$. For example, for the parameters of figures \ref{Fig:bifurcation_plot}-\ref{Fig:tau_WP2} we have $w=0.3166 z_0$ whereas $z_\mathrm{c}\approx 0.3461 z_0$, i.e. the difference is around 10\%. This means that the second solution still dominates over the first solution not only within the dynamic tunnel barrier but also in its immediate neighbourhood. However, the probability density of the second solution is evanescent at larger distances and therefore does not induce there a density flow towards the opposite contact, in contrast to the first solution. These findings are in agreement with the numerical observations for the electronic density in the gap between contacts obtained by the time-dependent density functional theory (TDDFT) \cite{Ludwig2020} in the case of relatively large gaps ($\sim6$~nm). Thus our theory is able to contribute to the understanding of the pulse-induced tunneling dynamics of the electron density in vicinity of the nanocontacts. Furthermore, we will discuss below that in the strongly non-adiabatic regime, with large values of the Keldysh parameter, $z_\mathrm{c}$  can considerably exceed $w$ [cf. inset to figure \ref{Fig:figure_validity}(a)] increasing the role of the second solution, in particular for gaps $\lesssim1$~nm. In terms of finding optimal complex trajectories and corresponding probabilities, this situation is captured by the small-distance scenario, treated separately in the end of this section.

For the imaginary part of the action acquired along the optimal trajectory  we obtain from equation \eqref{Eq:ImS_solved} with $v=0$ and $\tau_0=0$:
\begin{equation}\label{Eq:S_tilde_case1}
  \mathrm{Im}\tilde{S}=\frac{F_0^2}{2 m \Gamma^3}\Big((\gamma_\mathrm{_{HCP}}^2+1)\Gamma\tau_e-\tan\Gamma\tau_e\Big)\;.
\end{equation}
With equation \eqref{Eq:tun_prob_def} and the solution of
equation \eqref{Eq:tau_e_result_case_1} for $\tau_e$, we get
the tunneling probability
\begin{equation}\label{Eq:P_case1}
  P_e=\exp\left[-\frac{F_0^2}{\hslash m
  \Gamma^3}\Big((\gamma_\mathrm{_{HCP}}^2+1)\arctan\gamma_\mathrm{_{HCP}}-\gamma_\mathrm{_{HCP}}\Big)\right]\;.
\end{equation}
This is the result obtained in
\cite{Popov2001,Popov2001_JETP,Popov2004} by the ITM, where the
problem of atomic ionization by an ultrashort light pulse was
considered. In contrast to these works, here we use an alternative formulation of the quasiclassical approach to tunneling
in time-dependent fields \cite{LandauLifshitz3,Ganichev2002,Moskalenko2000}. With that, firstly, all the intermediate steps have been consistently explained avoiding \textit{ad hoc} assumptions.
As we will see below, this is especially important for the following consideration of more realistic waveforms of the applied light pulses. Secondly, the details of the two main solution branches for the time-dependent, real-space quasiclassical wave function have been clarified,
 with important physical consequences highlighted below.

Let us discuss the applicability range and limit cases. The quasiclassical description used here is appropriate
only if condition
$\hslash\omega\ll\Delta E$ is fulfilled
for the frequencies $\omega$ belonging to the spectral content of the pulse. This implies here
\begin{equation}\label{Eq:validity1}
  \hslash\Gamma\ll\Delta E\;.
\end{equation}
Another applicability condition is given by a general
requirement that the tunneling probability remains small:
\begin{equation}\label{Eq:validity2}
  P_e\ll 1\;.
\end{equation}
Finally, in our derivation of equation \eqref{Eq:P_case1} we have assumed that  the distance travelled by the electron under the barrier $w$ does not exceed the distance between the contacts $d$. This
restricts possible values of the generalized Keldysh parameter from above by the condition
\begin{equation}\label{Eq:d_crossover}
  \frac{dF_0}{\Delta E}>G(\gamma_\mathrm{_{HCP}})\equiv\frac{1}{\gamma_\mathrm{_{HCP}}^2}\ln(1+\gamma_\mathrm{_{HCP}}^2)\;.
\end{equation}
Failure to fulfill this condition always means that the small-distance scenario is realized, with a different procedure to find the optimal complex trajectory, as we described in section \ref{Sec:conditions_optimal}. We will return to this scenario below.
The dependence $G(\gamma_\mathrm{_{HCP}})$ is shown in figure \ref{Fig:figure_validity}(a).
For a fixed value of $\gamma_\mathrm{_{HCP}}$, it follows that $d$ should exceed $ G(\gamma_\mathrm{_{HCP}})\, \Delta E/F_0$.  For example, we have $G(\gamma_\mathrm{_{HCP}})=1/2$ if $\gamma_\mathrm{_{HCP}}\approx 1.585$. This implies $d>\Delta E/(2F_0)$: During the motion inside the classically forbidden region the electron absorbs an energy being equal to $\Delta E/2$. We will revisit conditions  \eqref{Eq:validity1}-\eqref{Eq:d_crossover} in discussion section, deliberating on realistic system parameters.

In the direct tunneling limit, i.e. $\gamma_\mathrm{_{HCP}}\rightarrow 0$, we have $w=\Delta E/F_0$. No energy is absorbed in the process.
Inequality \eqref{Eq:d_crossover} is satisfied when the barrier height $\Delta E$ does not exceed
the potential drop across the intercontact region corresponding to the peak value of the electric field of the light pulse, i.e. $d>\Delta E/F_0$.
Then decomposing the left hand side of equation \eqref{Eq:tau_e_result_case_1} in Taylor series we get $\Gamma\tau_e\approx\gamma_\mathrm{_{HCP}}$ and therefore $\tau_e\rightarrow 0$ with $\gamma_\mathrm{_{HCP}}\rightarrow 0$.
In the leading order in $\gamma_\mathrm{_{HCP}}$ in the exponent,
equation \eqref{Eq:P_case1} simplifies to
\begin{equation}\label{Eq:P_case1_small_gamma}
  P_e=\exp\left[-\frac{F_0^2}{\hslash m
  \Gamma^3}\frac{2}{3}\gamma_\mathrm{_{HCP}}^3\right]=\exp\left[-\frac{4}{3}\frac{\sqrt{2m}{(\Delta E)^{3/2}}}{\hslash F_0}\right]\;,
\end{equation}
which is just the well-known expression for the quasiclassical probability of
tunneling through a static triangular barrier. From equation \eqref{Eq:P_case1_small_gamma} and
equation \eqref{Eq:validity2} follows a condition limiting the
electric field strength \footnote{In the case of atomic ionization
this corresponds to the requirement that the maximum applied
electric field should be still much smaller than the
characteristic atomic electric field.}:
\begin{equation}\label{Eq:F0_limitation}
  F_0\ll \frac{\sqrt{2m \Delta E}\Delta E}{\hslash}\;.
\end{equation}
For $\gamma_\mathrm{_{HCP}}\gtrsim 1$, the applicability of the quasiclassical
approach would break down for smaller values of $F_0$ than dictated by
equation \eqref{Eq:F0_limitation}.

In the multiphoton limit, with  $\gamma_\mathrm{_{HCP}}\rightarrow \infty$, formally we get $\Gamma\tau_e \rightarrow \pi/2$  and equation \eqref{Eq:P_case1} reduces to
\begin{equation}\label{Eq:P_large_gamma}
  P_e=\exp\left[-\frac{\Delta E}{\hslash\, \pi^{-1}\Gamma}\right],
\end{equation}
where $\Delta E/(\hslash\, \pi^{-1}\Gamma)$ is the average number of absorbed photons in the multiphoton transition. Notice that
in contrast to the conventional continuous wave case here photons of various energies belonging to the pulse spectrum participate in the process. However, it is immediately clear that equation \eqref{Eq:P_large_gamma} cannot represent the correct limit case result because this limit can be reached by decreasing the electric field amplitude and keeping other system parameters constant. Obviously, when the electric field vanishes the probability must also vanish that is, however, not the case for equation \eqref{Eq:P_large_gamma}. In fact, as we discuss in \ref{App:paths}, the utilized quasiclassical description becomes invalid for too large values  the (generalized) Keldysh parameter.  The validity is restricted by the condition
\begin{equation}\label{Eq:P_large_gamma_validity_threshold}
  \gamma_{_\mathrm{HCP}} \lesssim \frac{\Delta E}{\hslash\, \pi^{-1}\Gamma}\;.
\end{equation}

In order to obtain the full picture, it is important to calculate the optimal complex trajectories and corresponding probabilities also in the small-distance scenario. Looking at equations \eqref{Eq:case2conditionA}-\eqref{Eq:case2conditionA}, we can eliminate there variables  $\mathrm{Re}v$ and  $\mathrm{Im}v$  by expressing
\begin{equation}\label{Eq:Rev}
  \mathrm{Re}v=-z_0 \Gamma s_1 (\tau_0, \tau_e)\;,
\end{equation}
from equation \eqref{Eq:case2conditionC} and
\begin{equation}\label{Eq:Imv}
  \mathrm{Im}v=\sqrt{\frac{2}{m}\Delta E}-z_0 \Gamma s_2(\tau_0, \tau_e)\;.
\end{equation}
from equation \eqref{Eq:case2conditionD}. Then we can find from equations \eqref{Eq:Nut} and \eqref{Eq:Zt}
\begin{equation}\label{Eq:Nu_tE}
  \mathcal{V} (t_E) = \frac{F_0}{m \Gamma} \tanh (\Gamma t_E)\;,
\end{equation}
\begin{align}\label{Eq:Z_tE}
  \mathcal{Z} (t_E, \tau_0 + i \tau_e) = \frac{F_0}{m \Gamma^2} \Big[&\ln \cosh \Gamma t_E - \frac{1}{2} \ln f(\Gamma \tau_0, \Gamma \tau_e) \nonumber\\
  & - i \, \arg \left(\cos\Gamma \tau_e \cosh\Gamma \tau_0 + i \, \sin\Gamma \tau_e \sinh \Gamma \tau_0\right)\Big]\;
\end{align}
that by inserting into equations  \eqref{Eq:case2conditionA}, \eqref{Eq:case2conditionB} and \eqref{Eq:case2conditionE} leaves us with  a system of three equations for  three variables: $t_E, \tau_0, \tau_e$. Solving this system we can determine these variables and find then the corresponding probability.

However, we can notice that equation \eqref{Eq:case2conditionE}
actually means that $\mathrm{Im} v=0$ or/and $\mathrm{Re} v + \mathrm{Re} \mathcal{V} (t_E)=0$
must be fulfilled. The first option would lead actually again to equations \eqref{Eq:first_tau_0_tau_e} and \eqref{Eq:second_tau_tau_e}
and then to equation \eqref{Eq:tau_e_result_case_1}, obtained above for the case when the electron leaves the classically forbidden region before reaching the opposite contact, that would contradict to the assumption of the small-distance scenario.
Choosing the option $\mathrm{Re} v + \mathrm{Re} \mathcal{V} (t_E)=0$ and using equations \eqref{Eq:Nu_tE} and \eqref{Eq:Z_tE}, equations \eqref{Eq:case2conditionA}, \eqref{Eq:case2conditionB} and \eqref{Eq:case2conditionE} can be  recast as
\begin{equation}\label{Eq:system_case2_eq1}
  \Gamma t_E = \arctanh s_1 (\tau_e)\;,
\end{equation}
\begin{align}\label{Eq:system_case2_eq2}
  \left[\Gamma \tau_0 - \arctanh s_1 (\tau_e)\right] s_1 (\tau_e) &+ \Gamma \tau_e \left[\gamma_\mathrm{_{HCP}} - s_2 (\tau_e)\right]
  \nonumber \\
  &
  +  \ln \cosh \arctanh s_1 (\tau_e) - \frac{1}{2} \ln f(\Gamma \tau_0, \Gamma \tau_e) = \frac{d}{z_0}\;,
\end{align}
\begin{align}\label{Eq:system_case2_eq3}
  \Gamma \tau_e s_1 (\tau_e) &+ \left[\arctanh s_1 (\tau_e)-\Gamma \tau_0 \right] \left[\gamma_\mathrm{_{HCP}} - s_2 (\tau_e)\right] \nonumber\\
  &- \arg \bigl(\cos\Gamma \tau_e \cosh\Gamma \tau_0 + i \, \sin\Gamma \tau_e \sinh \Gamma \tau_0\bigr)= 0\;.
\end{align}
In general the system of these three equations has to be solved numerically but we can notice that there is always a solution with $\tau_0 = 0$, leading also to $t_E =0$. This results in
\begin{equation}\label{case2_equation_12}
  \Gamma \tau_e \left[\gamma_\mathrm{_{HCP}} - \tan(\Gamma \tau_e)\right] - \ln \cos (\Gamma \tau_e) = \frac{d}{z_0}\;,
\end{equation}
from which we can find then also $\tau_e$. Note that this equation actually coincides with equation \eqref{Eq:t0_z_t_sech} if we take there $z=d$ and $t=0$.
Thus we have actually already analyzed its solutions and illustrated them  in figure \ref{Fig:bifurcation_plot} for a particular choice of parameters. For $d<z_\mathrm{c}$ there are two solution branches, whereas it is actually the second solution which plays the dominating role in this scenario.
Moreover,
for $w<d<z_\mathrm{c}$ where the first solution possesses also the optimal complex trajectory exiting into the classically allowed region before reaching the opposite contact and merging there with the real classical trajectory $z_\mathrm{re}(t)$, the second solution, having the optimal complex trajectory without this property, still leads to higher values of probability. Other possible solutions of equations \eqref{Eq:system_case2_eq1}-\eqref{Eq:system_case2_eq3}, which can be found numerically and cannot be related to our first and second solution branches, lead to much smaller probabilities than the dominating solution branch and may be therefore neglected.

Finally, using equation \eqref{Eq:ImS_solved_case2} we can find
\begin{equation}
  \mathrm{Im} \tilde{S}= \frac{F_0^2}{2 m \Gamma^3}\left[\Gamma \tau_e \left[\left\{\gamma_\mathrm{_{HCP}} - \tan(\Gamma \tau_e)\right\}^2 + \gamma_\mathrm{_{HCP}}^2 + 1\right] - \tan(\Gamma \tau_e)\right]\;.
\end{equation}
so that the resulting tunneling probability in the small-distance scenario can be expressed as
\begin{equation}
  P_e=\exp\left\{-\frac{F_0^2}{\hslash m
  \Gamma^3}\Big[\Gamma \tau_e \big(\gamma_\mathrm{_{HCP}} - \tan\Gamma \tau_e)^2 + \gamma_\mathrm{_{HCP}}^2 + 1\big] - \tan(\Gamma \tau_e)\Big]\right\}\;.
\end{equation}

Dependence of the tunneling probability on the Keldysh parameter is illustrated in figure \ref{Fig:figure_validity}(b) for the studied half-cycle pulsed driving in comparison with the driving by a FCP with a Gaussian envelope, continuous wave (CW) driving as well as the result that follows from equation \eqref{Eq:P_case1_small_gamma} (direct tunneling limit). The validity threshold following from equation \eqref{Eq:P_large_gamma_validity_threshold} is indicated by the gray bar labeled by ''v''. Note that for large electric-field strengths, and correspondingly small values of $\gamma$, the results for the tunneling probability converge to the direct tunneling limit in all cases. On the double logarithmic scale chosen for figure \ref{Fig:figure_validity}(b) it leads to an exponential behaviour $\log_{10} P=-ae^{\ln\!10\,\log_{10}\!\gamma}$ ($a>0$).
This agrees with the corresponding limit case result given in \cite{Reiss1980} (cf. p. 1797 there) but seems to disagree with  some later numerical calculations \cite{Bormann2010,Garg2020} used to model related experimental data and showing linear behaviour on the double logarithmic scale, $\log_{10} \!P=-a'-b'\log_{10}\!\gamma$  ($a',b'>0$), in the limit $\gamma\rightarrow 0$.
However, the latter calculations are actually based on the Reiss theory \cite{Reiss1980} so it remains unclear to us why they fail to reproduce the result of the direct tunneling limit inherent to the Reiss theory (or generally Keldysh-Faisal-Reiss theory) as mentioned above. The experimental data provided in \cite{Bormann2010} would not allow to differentiate strictly between the two dependencies. Concerning the experimental and theoretical data in  \cite{Garg2020} we do not recognize the underlying physical reason for a quite abrupt slope change at $\gamma\sim 1$ indicated there and cannot confirm it by our theory.

In discussion of the behaviour of the probability for large values of $\gamma_\mathrm{HCP}$ the crucial aspect is to know the value of the intercontact distance $d$. For very large $d$ the probability is always determined by the first solution and decays rapidly with increase of the Keldysh parameter, though somewhat slower than for the CW driving and even more slower than given by the direct-tunneling expression. In this consideration we have to keep in mind the restriction imposed by the condition \eqref{Eq:P_large_gamma_validity_threshold}. However, due to the decrease of the probability to very small values the discussion of the first solution for extremely large values of $\gamma_\mathrm{HCP}$ has anyway no practical sense. Moreover, for $d\lesssim 1~$nm it is the second solution that plays the dominating role with much larger probabilities for large $\gamma_\mathrm{HCP}$, not the first solution. At finite $d$ the second solution has a clear physical limit case for vanishing field strengths, i.e. for $\gamma_\mathrm{HCP}\rightarrow \infty$ in the context of  figure \ref{Fig:figure_validity}(b). This limit case corresponds to tunneling via a static rectangular potential barrier of height $\Delta E$ and width $d$, as we also illustrate in figure \ref{Fig:figure_validity}(b) for $d=0.6~$nm. We see that at such small gaps the second solution overtakes the first one already at values $\gamma_\mathrm{HCP}$ below 1, determined by  the condition $z_\mathrm{c}(\gamma_\mathrm{HCP})=d$, and leads then to much higher probabilities than the first solution for larger values of the Keldysh parameter. In result, the whole dependence of the tunneling probability on the field strength, converted to the Keldysh parameter, is significantly different from the result taking into account  the first solution only or equivalently the assumption of large $d$. This has to be kept in mind when discussing related experiment results similar to whose presented in \cite{Bormann2010} and \cite{Garg2020}.

\subsection{Realistic few-cycle pulse}\label{Sec:Realistic_Pulse}
To model a realistic FCP \cite{Brida2014} we use a wave packet with a rectangular spectral shape \cite{Moskalenko2015} and a variable flat phase.
The temporal profile determined by
\begin{equation}\label{Eq:Ft_fc}
  F(t)=\!\int_{\omega_1}^{\omega_2} \! F(\omega) e^{i\omega t}\mathrm{d}\omega+\mathrm{c.c.},
\end{equation}
where the Fourier transform $F(\omega)$ is constant and can be decomposed into a real positive amplitude $F_0$ and a phase factor factor $e^{i\phi}$:
\begin{equation}\label{Eq:Fw_fc}
  F(\omega)=\frac{F_0}{2\Delta\omega}e^{i\phi}.
\end{equation}
Here $\Delta\omega=\omega_2-\omega_1$ is the spectral bandwidth of the pulse and $\phi$ is the CEP.

In this part we limit our consideration to the optimal complex classical trajectory and tunneling probability, keeping in mind the structure of the quasiclassical wave function  discussed in the preceding subsection. Using equations \eqref{Eq:Ft_fc} and \eqref{Eq:Fw_fc} in  equations \eqref{Eq:Dv} and \eqref{Eq:Dz} we get
\begin{align}
  &\mathcal{V}(t')=\frac{F_0}{m\Delta\omega}\int_{\omega_1}^{\omega_2}\!\!\mathrm{d}\omega\, \frac{1}{\omega}\left[\sin(\omega t'+\phi)-\sin\phi\right],\label{Eq:Nut_fc}\\
  &\mathcal{Z}(t',t_0)=\frac{F_0}{m\Delta\omega}\int_{\omega_1}^{\omega_2}\!\!\mathrm{d}\omega\, \frac{1}{\omega}\left[\frac{1}{\omega}\cos(\omega t_0+\phi)-\frac{1}{\omega}\cos(\omega t'+\phi)+(t_0-t')\sin\phi\right].\label{Eq:Zt_fc}
\end{align}
Substituting these expressions into equations \eqref{Eq:case1conditionA}-\eqref{Eq:case1conditionC} and eliminating the variable $v$ leads to
the following equation system for the times $\tau_0$ and $\tau_e$:
\begin{align}
  &\int_{\omega_1}^{\omega_2}\!\!\mathrm{d}\omega\,  \sin(\omega\tau_0+\phi) i_1(\omega\tau_e)=0, \label{Eq:system_times_FCP1}\\
  &\tau_e\int_{\omega_1}^{\omega_2}\!\!\mathrm{d}\omega\,  \cos(\omega\tau_0+\phi) i_0(\omega\tau_e)=\gamma_{_\mathrm{FCP}} \label{Eq:system_times_FCP2}\;,
\end{align}
with
\begin{equation}\label{Eq:gamma_Keldysh_FCP}
   \gamma_{_\mathrm{FCP}}=\frac{\Delta\omega}{F_0}\sqrt{2m\Delta E}
\end{equation}
and $i_n(x)$ denoting the modified spherical Bessel functions of the first kind \cite{Abramowitz_book}. For each pair $\tau_0$ and $\tau_e$ we find the optimal complex trajectory and the tunneling probability determined by
\begin{equation}\label{Eq:Im_S_fc}
   \frac{2}{\hslash}\mathrm{Im}\tilde{S}=
 \frac{F_0^2}{\hslash m \Delta\omega^3}
  \left\{\gamma_{_\mathrm{FCP}}^2 \Delta\omega\tau_e
  +\Delta\omega\int_0^{\tau_e}\left[s_1^2(\tau)-s_2^2(\tau)\right]\mathrm{d}\tau\right\},
\end{equation}
where
\begin{align}
  &s_1(\tau)=\int_{\omega_1}^{\omega_2}\!\!\mathrm{d}\omega\,  \frac{1}{\omega} \sin(\omega\tau_0+\phi)
  \left[\cosh\omega\tau-\cosh\omega\tau_e\right], \label{Eq:s1_fc}\\
  &s_2(\tau)=\int_{\omega_1}^{\omega_2}\!\!\mathrm{d}\omega\,  \frac{1}{\omega} \cos(\omega\tau_0+\phi)
  \sinh\omega\tau\;. \label{Eq:s2_fc}
\end{align}

\begin{figure}
   \centering
  \includegraphics[width=15cm]{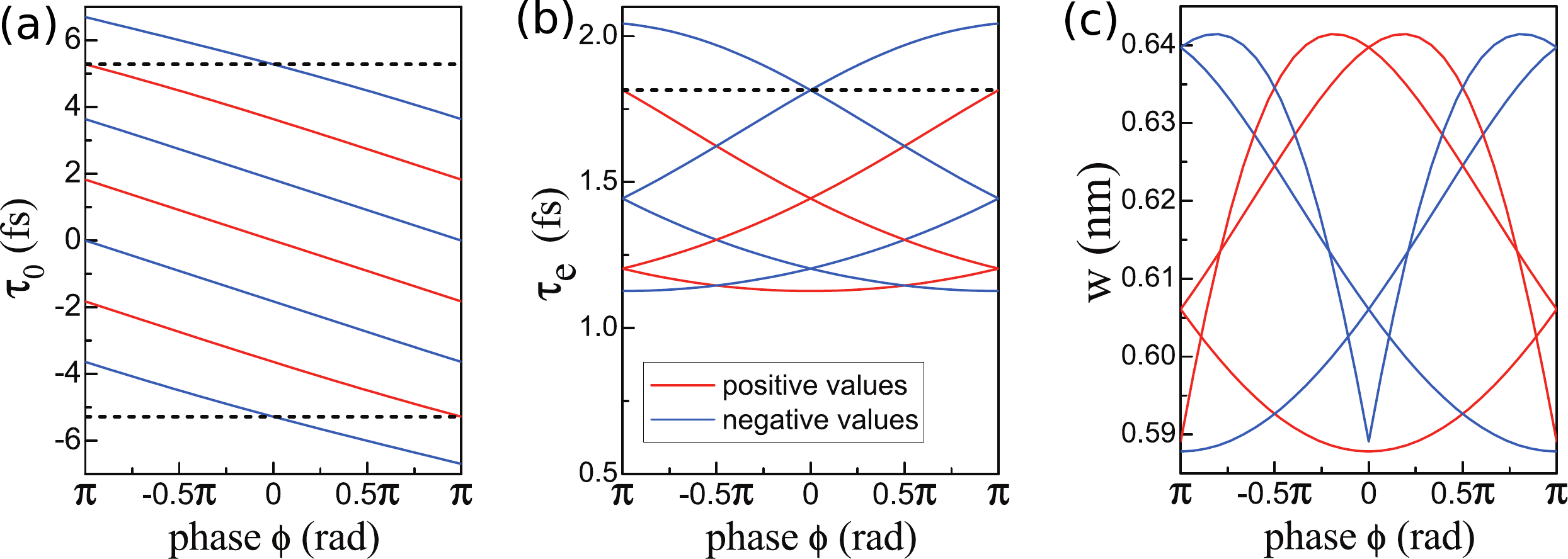}\\
  \caption{Dependence of the complex tunneling time $t_0$ and the tunneling  distance $w$ travelled by the electron under the barrier (when it does not exceed the distance between the contacts $d$) on the carrier-envelope phase (CEP) $\phi$ of the driving few-cycle pulse (FCP). (a) shows multiple solutions for $\tau_0=\mathrm{Re}t_0$ corresponding to the field peaks around the FCP center. Red (blue) color is used when the tunneling direction is from the left (right) contact towards the right (left) contact, determined by the corresponding sign of the electric field. Dashed lines limit the region, where we always have exactly 6 solutions being closest to the center of the applied pulse, independent of the CEP value. (b) illustrates the  behaviour of  corresponding $\tau_e=\mathrm{Im}t_0$, whereas the lines show the absolute values. The sign, determined by the tunneling direction, is reflected in the line color.
  $\tau_e$ increases with the separation between the corresponding $\tau_0$ and the pulse center. Dashed line again limits the region with 6 most relevant solutions. (c) depicts the the tunneling  distance $w$ for these solutions. FCP parameters: $F_0=4$~{eV/nm}, $\omega_1/(2\pi)=180$~THz, $\omega_2/(2\pi)=330$~THz; other parameters: $\Delta E=5$~eV.}
  \label{Fig:tau_both}
\end{figure}

\begin{figure}
   \centering
  \includegraphics[width=15cm]{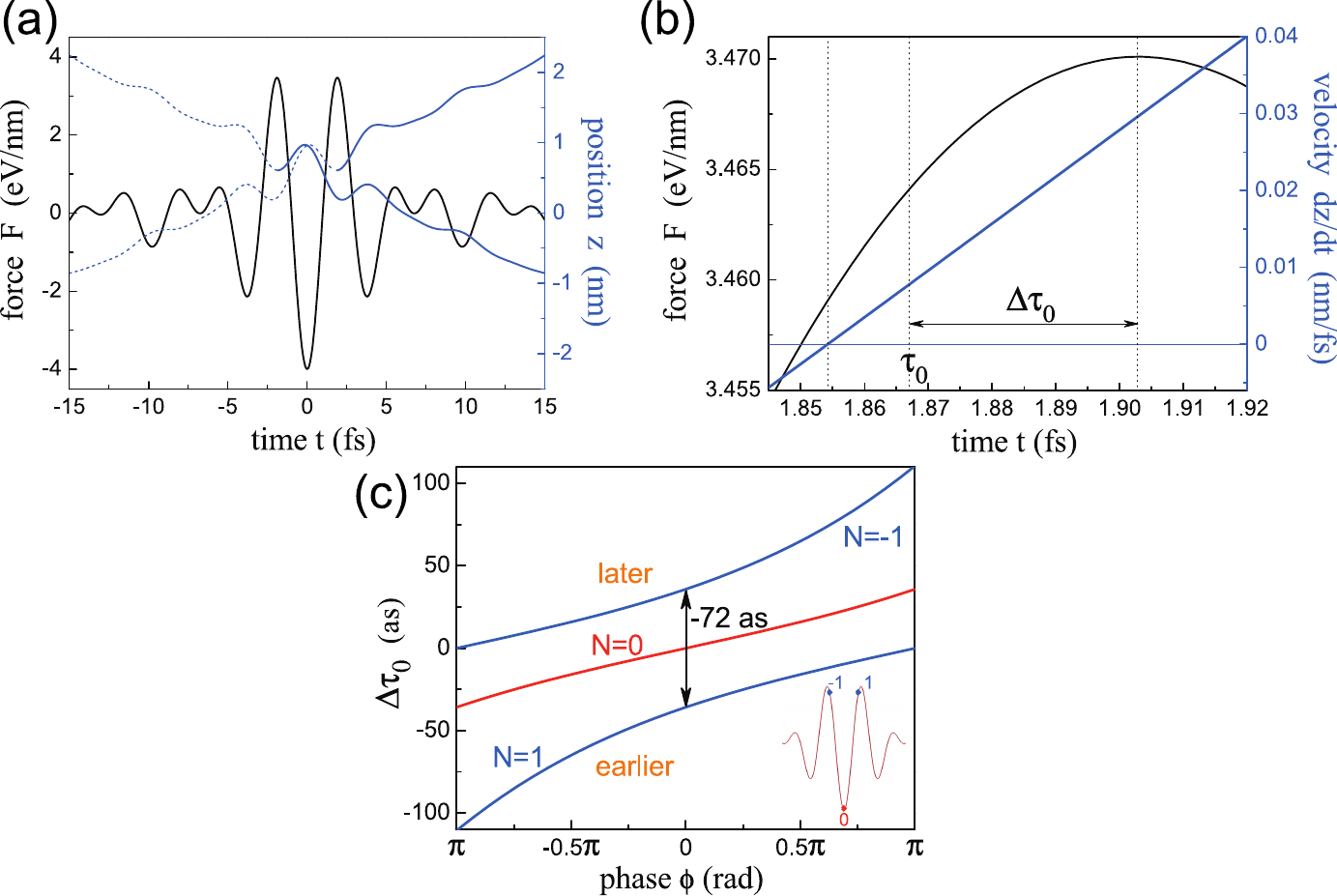}\\
  \caption{(a) Temporal profile of the light-induced force $F(t)=e\mathcal{E}(t)$ (black line) determined by the applied electric field pulse $\mathcal{E}(t)$, shown for the same parameters as in figure \ref{Fig:tau_both} and $\phi=0$. The same plot illustrates the classical trajectories $z_\mathrm{re}(t)$ originating from the two main positive peaks. Full (dashed) blue lines are for $t>\tau_0$ ($t<\tau_0$), with the fictitious reflection happening around $t=\tau_0$ and the contact boundaries, interrupting the trajectories, being ignored for this plot. (b) Zoomed region in vicinity of the first positive peak. Blue line shows now the velocity of the fictitious/real electron. Neither $\tau_0$  nor the time moment when this velocity vanishes coincide with the temporal position of the peak.  Difference between $\tau_0$ and the latter is denoted as $\Delta\tau_0$. (c) Dependence of $\Delta\tau_0$ on the CEP $\phi$ for 3 main peaks of the pulse. Here $N$ numbers the peaks according to the inset.}
  \label{Fig:atto_delay}
\end{figure}

The behaviour of the solutions of equations \eqref{Eq:system_times_FCP1} and \eqref{Eq:system_times_FCP2} in dependence on the CEP is illustrated in figures \ref{Fig:tau_both}(a) and \ref{Fig:tau_both}(b)  [for the temporal shape of the chosen FCP, see figure \ref{Fig:atto_delay}(a)]. We see multiple solutions in these figures. Each solution can be actually attributed to a peak of the FCP. Tunneling takes place at a time moment close to the temporal position of the corresponding peak, in the sense that there is a value of $\tau_0$ being close to that position (but generally not exactly coinciding with it, as we will discuss below). The absolute value of the imaginary tunneling time $\tau_e$ increases monotonically with the distance of the peak from the center of the pulse as it is changing with $\phi$. In figure \ref{Fig:tau_both}(c) for each of the solutions we show the tunneling  distance $w$ travelled by the electron under the barrier, whereas for the case when this quantity is smaller than the distance between the contacts. We see that  $w$ behaves non-monotonically in dependence on the phase. For 6 main solutions, having the smallest values of $|\tau_0|$ and $|\tau_e|$ with respect to the remaining solutions, the values of $w$ are confined to a relatively narrow interval, which for the selected parameters constitutes around 0.05~nm.

For a FCP the distance between contacts $d$ with respect to $w$ brings an additional aspect for the resulting charge transfer. When $d$ is large enough with respect to $w$ the electron might have no chance to reach the opposite contact after the tunneling through the barrier because the electric field changes its polarity after a certain time and might drive the electron back to the original contact. The corresponding motion is determined by the corresponding classical trajectory $z_\mathrm{re}(t)$. In figure \ref{Fig:atto_delay}(a), as an example, we show two such trajectories originating from two positive peaks being closest to the center of the pulse at  $\phi=\pi/2$. Whereas the trajectory belonging to the right peak does not return back to the left contact, the electron appearing in the gap around the left peak moves a certain distance towards the right contact but then is turned back due to the changed polarity of the field and flies back to the left contact. If the distance between the contacts is too large the latter electron never reaches the opposite contact. Then, neglecting possible reflections by the contact boundaries, it does not contribute to the resulting charge transfer. For larger gaps between the nanocontacts it is thus essential to model not only the tunneling process but also the following dynamics in the classically allowed region. An example can be found, e.g., in \cite{Ludwig2020}. Here we focus our attention on the regime when $d>w$ for any of the contributing trajectories but $d$ is still small enough so that the tunneling process determines the charge transfer. Before that let us look more closely at the positions of $\tau_0$ associated with the exit point of the standard complex classical trajectory to the classically allowed region.

In figure \ref{Fig:atto_delay}(b) we zoom into the time interval close to the position of  one of the field peaks. We can observe that the corresponding value of $\tau_0$ does not exactly coincide with the position of the peak. In the figure we also show the electron velocity found from the classical trajectory $z_\mathrm{re}(t)$. The time moment where the velocity vanishes determines the turning point of the trajectory. We see that this time moment generally does not coincide neither with $\tau_0$ nor with the peak-field position.
This fact is remarkable in view of recent suggestions to associate the tunnel exit time with the time moment when the velocity vanishes that can be determined, e.g., by the backpropagation of the emitted electronic wave packet \cite{Ni2016,Ni2018,Rost2020}. One justification argument behind these suggestions is that ``the pure tunnelling dynamics in a semi-classical description is characterized by an imaginary momentum component in at least one degree of freedom'' whereas ``all momenta are real in classically allowed regions''. However, in general the \textit{ad hoc} assumption that the momentum of the electron during its underbarrier dynamics has a vanishing real part, implied by the ITM, has no justified reason to hold. As it is in fact happens for the situation of figure \ref{Fig:atto_delay}(b), this momentum runs generally over \textit{complex} and not  \textit{purely imaginary} values. Thus at the moment when the electron enters the classically allowed region and its momentum becomes real, it should not necessarily also vanish. In fact, it depends on the particular chosen path in the complex time plane. For the standard vertical path, when $t$ changes from $t_0$ to $\mathrm{Re}(t_0)=\tau_0$, considering the situation of figure \ref{Fig:atto_delay}(b) we can see that  the momentum does not vanish at $t=\tau_0$ but acquires a finite real value. The momentum, of course, vanishes if an alternative, non-vertical path reaching the real time axis at $t$ with $\dot{z}_\mathrm{re}(t)=0$  is selected.   The discussion on the definition and meaning of the tunneling time with the related controversy \cite[and references therein]{Landsman2015,Rost2020,Kheifets2020}
is out of the scope of the present paper but we hope that our findings  bring an additional useful insight in this context.

It is interesting to analyze the attosecond time shift  $\Delta\tau_0$ between $\tau_0$ and the temporal position of the corresponding peak of the electric field. We illustrate this in  \ref{Fig:atto_delay}(c). For $\phi=0$ there is no shift for the central negative [in terms of the induced force $F(t)=e\mathcal{E}(t)$] peak, whereas the electron is emitted later (earlier) when it might be expected for the proceeding (succeeding) peak. When the CEP $\phi$ is changed away from zero  the time shift $\Delta\tau_0$  appears also for the central peak, due to the break of symmetry. Generally, the absolute value of $\Delta\tau_0$  grows as the position of the peak moves away from the center of the FCP, with its sign being positive (negative) for $\tau_0<0$ ($\tau_0>0$). It would be interesting to find if this effect is possible to measure experimentally.

\begin{figure}
   \centering
  \includegraphics[width=14cm]{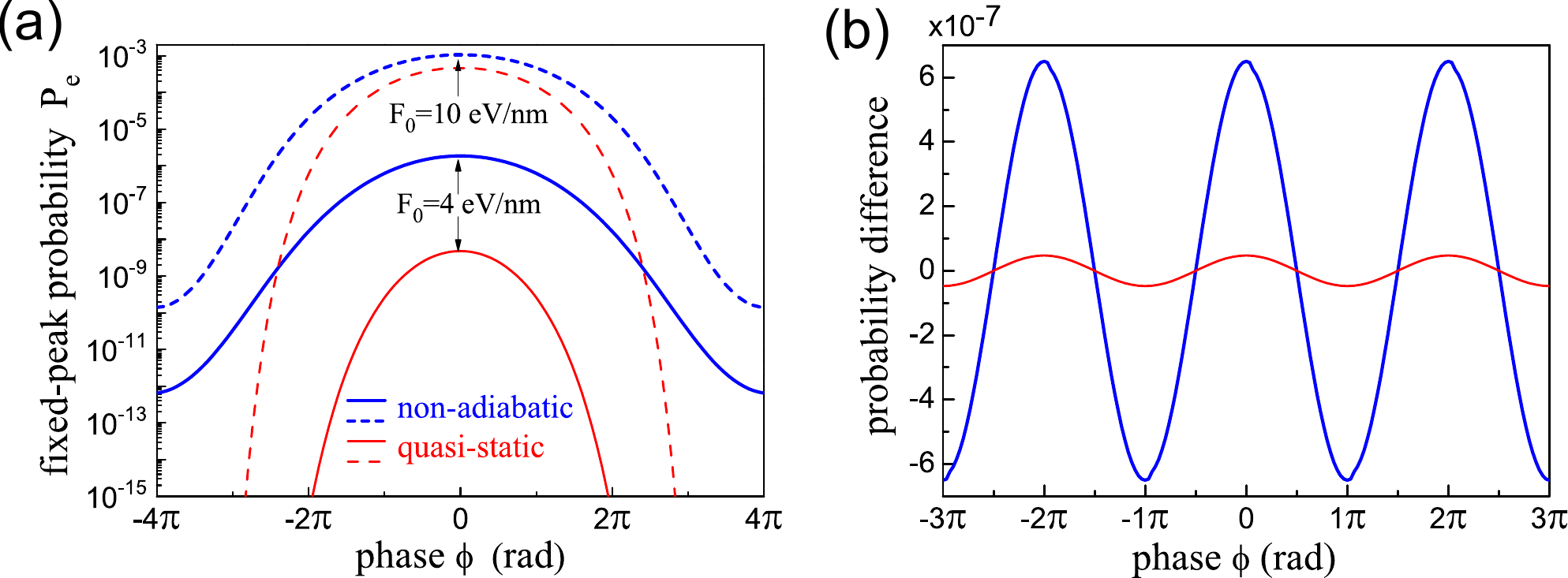}\\
  \caption{(a) Dependence of the tunneling probability corresponding to a fixed FCP peak on the CEP $\phi$ for two selected values of the pulse amplitude $F_0=e\mathcal{E}_0$. Other parameters are as in figure \ref{Fig:tau_both}. The result of the non-adiabatic approach is compared to the quasi-static approximation. (b) Difference in probabilities resulting from the whole waveform of the FCP. Here we took $F_0=4$~eV/nm and $d=0.7$~nm. Color scheme as in (a).}
  \label{Fig:probability}
\end{figure}

Finally, having obtained the solutions for $t_0$ based on equation \eqref{Eq:Im_S_fc} we can calculate the resulting transition probabilities belonging to each of these solutions. In order to represent the result for all main solutions simultaneously it is convenient to fix a particular peak of the electric field and the related solution and then evaluate the dependence of the tunneling probability on the phase in the extended phase scheme, where the CEP $\phi$ can take any real values, not limited by the interval $(-\pi,\pi)$. The result is shown in figure \ref{Fig:probability}(a) for two different values of the peak electric-field amplitude. For comparison we also plot the results corresponding to the quasi-static approximation when the probability is calculated using the direct-tunneling approach with the static potential determined by the value of the electric field at the corresponding peak. We see that the quasi-static approximation significantly underestimate the probability, especially for lower field values. For lower $F_0$ the effect is pronounced already at $\phi=0$ (central peak). The non-adiabatic enhancement grows with increase of the absolute value of the CEP, as the distance between the peak and the FCP centre raises. Therefore, in order to determine the charge transfer from the whole FCP it is sufficient to take into account just several contributions coming from the closest peaks to the FCP centre, with the respective sign determined by the direction of the field.

In order to calculate the resulting probability difference between positive and negative contributions we limited the consideration to the six closes peaks to the FCP centre, with $\tau_0$ and $\tau_e$ limited to the intervals indicated in figures  \ref{Fig:atto_delay}(a) and \ref{Fig:atto_delay}(b). For each electron emerging from the tunneling region we calculate its classical trajectory and take its contribution into account only if it reaches the opposite contact avoiding  in the meantime the original contact. For larger values of the distance between the contacts $d$ it happens that the classical field-induced dynamics plays a major role in determining the overall charge transfer \cite{Ludwig2020}. It is then insufficient to limit the consideration to optimal complex trajectories $z_\mathrm{opt}(t)$ and related real trajectories $z_\mathrm{re}(t)$. All complex trajectories (or in other words whole emitted electronic wave packets) emerging at each field peak should be taken into account. It can happen that the part of the wave packet reaching the opposite contact does not contain the optimal trajectory at all. In order to have a situation dominated by the tunneling process we consider a configuration when $d$ is close to the tunneling distance $w$ for all real trajectories but still always exceeds it by a certain amount, so that we may consider only the wave-packet-like (first) solution (cf. section \ref{Sec:Ideal_HCP}) neglecting the evanescent (second) solution. In this regime we can also neglect the effect of interference between electronic wave packets  emitted at neighbouring electric-field peaks \cite{Popov2004,Popruzhenko2014}. We take the parameters of figure \ref{Fig:tau_both}(c) and select $d=0.75$~nm.

The resulting probability difference is shown in figure \ref{Fig:probability}(b). For comparison we show also the results for the quasi-static approximation when the electron would not gain energy during tunneling. As expected, we see that the non-adiabatic description leads to significantly higher values of the probability difference than the  na\"{\i}ve quasi-static approach. We can also nicely observe the modulation of our calculated quantity determining the total charge transfer induced by the FCP in dependence on the CEP. Thus the direction of the charge transfer can be controlled by FCP on an ultrashort time scale, in agreement with the
experimental observations and other theory predictions \cite{Rybka2016,Ludwig2020,Ludwig2020_2,Garg2020}. In the studied case the modulation can be fitted by a cosine function so that the difference of the fit to the calculation result in figure \ref{Fig:probability}(b) is visually barely distinguishable and we therefore did not plot it separately.


\section{Discussion of the relevant system parameters and approximations}

Let us at first briefly discuss the relevant  parameters of light pulses (cf., e.g, reference \cite{Brida2014}). Taking typical pulses in the near-infrared with wavelength $\lambda\sim 1.1~\mu$m we have the half-cycle duration $\sim 2$~fs that gives the relevant scale for $1/\Gamma$ in case of the considered ideal HCPs.  Realistic FCPs can be as short as  $\tau_{_{\mathrm{FWHM}}}\sim6$~fs. The peak electric-field strength $E_0$ can be estimated from the pulse energy $E_{\mathrm{p}}\sim200$~pJ that is focussed to an area of $A\sim20~\mu$m$^2$. Taking into account that the pulse energy can be expressed as $E_{\mathrm{p}}\approx\frac{c\epsilon_0}{2}|E_0|^2A\tau_{_{\mathrm{FWHM}}}$, where $c$ denotes the speed of light and $\epsilon_0$ is the vacuum permittivity, we arrive at  $E_0\sim1$~V/nm. Between the nanocontacts the field is amplified due to the plasmonic enhancement by a factor $\theta$, which can take values up to $\sim100$ but it is strongly dependent on a particular realized configuration. The pulse typically also experiences certain phase shifts or/and distortion. In our modelling we operate with the anticipated field in the gap. In our illustrating examples, we orient ourselves on $E_0=4$~V/nm and $E_0=10$~V/nm inside the gap.

 Next, we want to review the parameters of the nanocontacts. We assume that they are made of gold having the Fermi energy $E_{_\mathrm{F}}\sim5.5$~eV.
 Depending on the manufacturing method, currently different sizes of the nanogap width $d$ are possible: from around 30~nm  \cite{Rybka2016} to 6~nm \cite{Ludwig2020,Ludwig2020_2} and further to subnanometer values in break junctions \cite{Stolz2014,Garg2020}. The height of the effective energy barrier is determined by the gap medium or it can be also influenced by the properties of the utilized substrate. For example, we would have  $\Delta E\approx 5.1$~eV for a Au/vacuum/Au composition of the nanocontacts and $\Delta E\approx 4.2$~eV for a Au/SiO$_2$/Au junction. In our calculations we took $\Delta E\approx 5$~eV.

Based on the above parameters of the driving light and nanocontacts we can estimate other important relations for our study. The relation of the average photon energy to the energy barrier width amounts to $\hslash\Gamma/\Delta E\sim0.07$ so that one of the conditions for the validity of the quasiclassic approach given by equation \eqref{Eq:validity1} can be seen as well satisfied. Concerning the (generalized) Keldysh parameter, assuming $E_0=4$~V/nm we estimate  $\gamma_{_\mathrm{HCP}}=0.94$.  This value corresponds to the intermediate regime between the direct tunneling and multiphoton ionization and we used it for many of our illustrations. For $E_0=10$~V/nm we have a 2.5 times smaller value of $\gamma_{_\mathrm{HCP}}$. The other validity condition given by equation \eqref{Eq:validity2} is also well satisfied for all parameter values discussed above.

Further, it is useful to review the characteristic spatial scales of the investigated system. The maximum tunneling distance for an electron at the Fermi level $w_\mathrm{F,max}=\Delta E/(|e|E_0)$ constitutes $\sim1.25~\mathrm{nm}$ taking $E_0=4$~V/nm. It decreases to just $\sim0.31~\mathrm{nm}$ for $E_0=10$~V/nm. For the electrons at the bottom of conduction band in the contacts the maximum underbarrier distance to overcome would be $\sim2.6$~nm ($\sim0.65$~nm) for $E_0=4$~V/nm ($E_0=10$~V/nm). The characteristic length $z_0$ defined by equation \eqref{Eq:z0} is estimated to $\sim2.8$~nm ($\sim0.7$~nm) for $E_0=10$~V/nm ($E_0=4$~V/nm) with $1/\Gamma=2$~fs.

Beyond the  evaluation of the tunneling probability for a single electron at the Fermi level we can roughly estimate the total number of electrons transferred between the nanocontacts by the applied pulse. We take the electron density of gold {$n_{_\mathrm{Au}}=5.9\times10^{28}~$m$^{-3}$}, assume that the area of the nanocontacts constitutes $\sim100$~nm$^2$,  calculate the Fermi velocity of the electrons $v_{_\mathrm{F}}$ from $E_{_\mathrm{F}}$ as $1.4\times10^6$~m/s, and estimate the number of electrons with energy close to the Fermi level and hitting the boundary in the temporal vicinity of the field peak as $(100~\mathrm{nm}^2)\times (2~\mathrm{fs}/10)\times(n_{_\mathrm{Au}}/20) \times v_{_\mathrm{F}}\sim 100$. Depending on the resulting probability difference being in the range $10^{-7}-10^{-3}$ this gives $2\times10^{-5}-2\times10^{-1}$ transferred electrons per pulse. With the standard pulse repetition rate of 40~MHz this amounts to $\sim10^{3}-10^{7}$ electrons per second or approximately $10^{-4}-1$~pA. Note that whereas our theory does not allow for a quantitatively precise evaluation of the magnitude of the transferred charge it provides its dependencies on various parameters of the system, and that in an analytical or semi-analytical way.

There is one effect neglected in our consideration that potentially can significantly influence the tunneling barrier and resulting probability. When the electron is outside the metal in a static case it should experience the interaction with its own image charge created inside the metal contact.
Taking the image charge into account leads to the modification of the triangular shape of the barrier leading to the so-called Schottky-Nordheim barrier \cite{Nordhiem1928,Forbes2007} given by
\begin{equation}\label{Eq:SN_barrier}
   U_{\mathrm{SN}}(z)=-eE_0z-\frac{e^2}{16\pi\epsilon_0z}\;,
\end{equation}
where the factor $16\pi$ appears here in place of the usual $4\pi$ for the usual Coulomb potential because the distance from the tunneling electron
to the image charge is the double of the distance to the metal, leading to the additional factor $1/4$ in the force and therefore also in the potential.
This leads to the lowering of the maximum height of the potential barrier with respect to $\Delta E$ by \cite{Schottky1914}
\begin{equation*}\label{Eq:Schottky}
   \Delta E_{\mathrm{S}}=\sqrt{\frac{e^3E_0}{4\pi\epsilon_0}}\;.
\end{equation*}
 For $E_0=4$~V/nm ($E_0=10$~V/nm) this amounts to  $\sim2.4~\mathrm{eV}$ ($\sim3.8~\mathrm{eV}$) reduction and could even remove the barrier completely for higher fields. However, we have to keep in mind that we deal here with a charge transfer process that takes place on an ultrashort time scale. In fact, the charge density reorganization leading to the appearance of the image charge interaction requires some time to be formed, determined by the inverse plasmon frequency. The image charge effect becomes essentially time-dependent and can be taken into account, e.g., by
using the corresponding velocity-dependent potential \cite{Ray1972,Jonson1980}. One can expect that the resulting barrier reduction is considerably lower than following from equation \eqref{Eq:SN_barrier} \cite{Ray1972,Jonson1980,Echenique1987,Mukhopadhyay1987}. We think that our quasiclassical approach, due to its time-dependent nature, has a good potential to be able to incorporate the image charge effect and to ultimately clarify the importance of the dynamic barrier reduction.

\section{Conclusion and outlook}
We have presented a quasiclassical theory for the description of  tunneling and charge transfer in nanocontacts that is driven by half-cycle and phase-controlled few-cycle pulses. The theory is capable to account for the dynamics of the underbarrier motion and the energy absorption taking place during this process. Based on a simple model of an ideal half-cycle we are able to construct analytical solutions for the main solution branches of the electronic wave function in the classically forbidden region as well as after exiting out of this region. We have derived the expression for the tunneling probability that was already known from the imaginary time method in the case of sufficiently large distances between the contacts but now it has been determined for any intercontact distances. We have found that for larger intercontact distances the solution branch corresponding to a rising electron density inside the barrier and forming an outgoing wave packet plays the dominating role in most of the classically allowed region whereas the solution branch with the falling electron density inside the barrier eventually forms an evanescent wave that can be neglected, unless we consider the behaviour in vicinity of the tunnel exit. Here the evanescent-wave solution already may start to give a larger probability. This is especially pronounced in the strongly non-adiabatic regime with higher values of the Keldysh parameter, where there is an extended spatial region where the evanescent-wave solution dominates over the solution corresponding to the outgoing wave packet with a classical trajectory.
It is one important finding of this work that this effect has to be taken into account if the boundary of the second contact occurs to be in this region.
At very short distances such that the electron stays always in the classically forbidden region the roles of the solution branches definitively interchange and the falling-density solution branch plays the leading role. Based on these results we were able to calculate the dependence of the tunneling probability on the strength of the applied field, also in the regime where a crossover between both solution branches takes place.

Further, in case of the phase-controlled few-cycle pulses we used our theory to find analytical solutions for the complex tunneling times and probabilities which determine the amount of the total induced charge transfer through the nanogap. In particular, we studied the configuration when the intercontact distance is such that the outgoing wavepacket solution with corresponding real trajectories can be used to calculate the resulting probabilities. We compared the turning points of these trajectories with the real parts of the tunneling times and with the temporal positions of the field peaks. We found that in general all these quantities are different, with temporal shifts being in the attosecond range for typical parameters. The amount of the transferred charge and  the transfer direction can be controlled by the carrier-envelope phase of the pulse, whereas the  values obtained in the non-adiabatic regime are significantly higher comparing to the quasi-static direct tunneling approximation.


One natural extension of the studied problem can be a consideration of a more complex spatial
structure of the tunneling region, e.g., having an additional quantum well inside the tunneling
barrier \cite{Kulkarni2015}.  In such a case one may expect an intriguing interplay between the tunneling process
steered by an ultrashort pulse and the energy level structure
inside the quantum well that should enable an ultrafast selective
population of the levels. In order to treat such a problem, a generalization of our method beyond the exponential accuracy \cite{Milosevich2006,Popruzhenko2014} is probably required, which also represents an important and interesting task by itself. Finally, an appropriate inclusion of the dynamical image effect \cite{Ray1972,Jonson1980,Echenique1987,Mukhopadhyay1987} into the description might lead to further improvement of our understanding of the light-induced tunneling in nanocontacts.

%


\section*{Acknowledgements} \addcontentsline{toc}{section}{Acknowledgements}

This research was supported by Basic Science Research Program through the National Research Foundation of Korea (NRF) funded by the Ministry of Education (2019R1A6A1A10073887). Funding by the DFG (SFB 767) is also gratefully acknowledged. We thank Daniele Brida,  Denis V. Seletskiy, Alfred  Leitenstorfer, and Irina N. Yassievich for helpful discussions.

\appendix

\section{Paths encircling singularities in the complex time plane}\label{App:paths}
For the pulse shape as in equation \eqref{Eq:soliton_shape}, $\mathcal{Z}(t',t_0)$, defined in equation \eqref{Eq:Dz} and entering equations determining $t_0$, is a multivalued function. The value depends on a particular integration path $\delta$ between $t_0$ and $t'$.
For real $t'$, we can calculate  $\mathcal{Z}(t',t_0)$ as a sum of an integral along the standard path $\delta_\mathrm{s}$ and an integral along the closed path $\mathring{\delta}=\delta_\mathrm{s}-\delta$, as discussed in the end of section \ref{Sec:action_wave_packet}. Proceeding in this way we write
\begin{equation}\label{Eq:Z_paths}
  \mathcal{Z}(t',t_0)=z_0\int_{\delta_\mathrm{s}}\tanh \tilde{t}''\mathrm{d}\tilde{t}''+z_0\oint_{\mathring{\delta}}\tanh \tilde{t}''\mathrm{d}\tilde{t}''.
\end{equation}
The integral in the second term can be evaluated using Cauchy's residue theorem and we get
\begin{equation}\label{Eq:Z_paths_Cauchy}
  \mathcal{Z}(t',t_0)=z_0\left[\int_{\delta_\mathrm{s}}\tanh \tilde{t}''\mathrm{d}\tilde{t}''+2\pi i N\right],
\end{equation}
where the total winding number
\begin{equation}\label{Eq:N}
  N=\sum_k N(\mathring{\delta},a_k)
\end{equation}
is a sum of all winding numbers $N(\mathring{\delta},a_k)$ of the path $\mathring{\delta}$ for each singularity point $a_k=\frac{\pi}{2}+k\pi$ ($k\in\mathbb{N}$). The remaining integral along $\delta_\mathrm{s}$ in equation \eqref{Eq:Z_paths_Cauchy} has two parts: one where $\tilde{t}''$ runs parallel to the imaginary axis from $(\Gamma\tau_0,\Gamma \tau_e)$ to  $(\Gamma\tau_0,0)$ and another from $(\Gamma\tau_0,0)$ to $(t',0)$. This leads to
\begin{equation}\label{Eq:Z_paths_Cauchy_par}
  \mathcal{Z}(t',t_0)=z_0\left[-i\int_0^{\Gamma\tau_e}\tanh (\Gamma\tau_0+i\tilde{\tau})\mathrm{d}\tilde{\tau}
  +\int_{\Gamma\tau_0}^{t'}\tanh (\tilde{\tau})\mathrm{d}\tilde{\tau}+2\pi i N\right].
\end{equation}
Evaluating the second integral in the brackets we obtain
\begin{equation}\label{Eq:Z_paths_Cauchy_res}
  \mathcal{Z}(t',t_0)=z_0\left[-i\int_0^{\Gamma\tau_e}\tanh (\Gamma\tau_0+i\tilde{\tau})\mathrm{d}\tilde{\tau}
  +\ln\cosh\Gamma t'-\ln\cosh \Gamma \tau_0+2\pi i N\right].
\end{equation}

It is instructive to limit our attention to the imaginary part of  $\mathcal{Z}(t',t_0)$ only, since it is sufficient when we want to determine $t_0$ in the case of optimal trajectories. From equation \eqref{Eq:Z_paths_Cauchy_res} we get
\begin{equation}\label{Eq:Z_paths_Cauchy_res_Im}
  \mathrm{Im}\mathcal{Z}(t',t_0)=-z_0\left[\mathrm{Arg}\, \eta(\Gamma\tau_0,\Gamma\tau_e)
  +2\pi n_\mathrm{tot}\right],
\end{equation}
where $\mathrm{Arg}\, \eta$ denotes the principal value of  the multi-valued  argument function $\mathrm{arg}\, \eta$ of a complex variable
\begin{equation}\label{Eq:eta}
    \eta(x,y)=a(x,y)+i b(x,y)
\end{equation}
with
\begin{eqnarray}
     a(x,y)&=&\cosh x\cos y,  \label{Eq:a}\\
     b(x,y)&=&\sinh x\sin y \label{Eq:b}.
\end{eqnarray}
 $\mathrm{Arg}\, \eta(x,y)\in(-\pi,\pi]$ can be calculated as
 \begin{equation}\label{Eq:Arg_eta_calc}
    \mathrm{Arg}\, \eta(x,y)=2 \arctan\left[\frac{b(x,y)}{|\eta(x,y)|+a(x,y)}\right],
 \end{equation}
 where $|\eta(x,y)|\equiv \sqrt{f(x,y)}$ (positive sign of the root is taken) and $f(x,y)=a^2(x,y)+b^2(x,y)$ is introduced in equation \eqref{Eq:f_HCP}.
 The integer number $n_\mathrm{tot}$ in equation \eqref{Eq:Z_paths_Cauchy_res_Im} is given by
 \begin{equation}\label{Eq:n_tot}
    n_\mathrm{tot}=n'-N,
 \end{equation}
 where $n'=[\Gamma \tau_e/(2\pi)]$ and  $[x]$ denotes here the nearest integer function of a real number $x$.

\begin{figure}[t!]
   \centering
  \includegraphics[width=10cm]{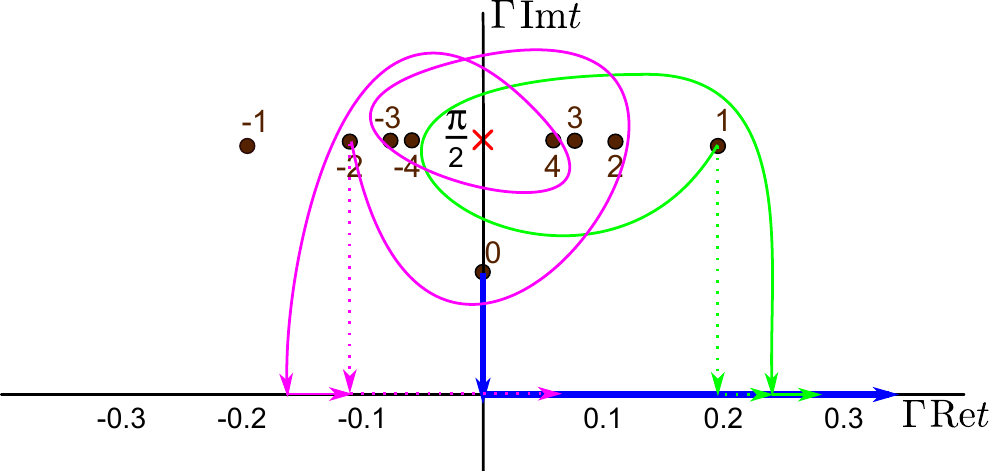}\\
  \caption{Solutions (full points) of the system \eqref{Eq:first_tau_0_tau_e_full},\eqref{Eq:second_tau_0_tau_e_full} determining the initial complex time $t_0=\tau_0+i\tau_e$ and corresponding representative paths (full color lines) in the complex time plane. The cross indicates a singularity point. The solutions and paths are labeled by
  $n_\mathrm{tot}$ defined in equation \eqref{Eq:n_tot},  having here the same absolute value as the winding number $N$ [cf. equation \eqref{Eq:N}] but the opposite sign. Solutions with
  $|n_\mathrm{tot}|> 4$ are not shown in this figure, just to keep a better overview. Dashed lines for $n_\mathrm{tot}=1$ and $n_\mathrm{tot}=-2$ indicate the  corresponding standard paths $\delta_\mathrm{s}$. Here we used  $\gamma_{_\mathrm{HCP}}=0.94$.}
  \label{Fig:paths_encircling}
\end{figure}

 From equations \eqref{Eq:case1conditionA}-\eqref{Eq:case1conditionC} we get then the following equation system for the determination of $\tau_e$ and $\tau_0$:
 \begin{align}
  &\Gamma\tau_e\frac{\sinh(2\Gamma\tau_0)}{2f(\Gamma\tau_0,\Gamma\tau_e)}=\mathrm{Arg}\, \eta(\Gamma\tau_0,\Gamma\tau_e)
  +2\pi n_\mathrm{tot}\;,
  \label{Eq:first_tau_0_tau_e_full}\\
  &\frac{\sin(2\Gamma\tau_e)}{2f(\Gamma\tau_0,\Gamma\tau_e)}=\gamma_\mathrm{_{HCP}}\;.
  \label{Eq:second_tau_0_tau_e_full}
\end{align}
In figure \ref{Fig:paths_encircling} we illustrate the behaviour of the solutions of this system in dependence on $n_\mathrm{tot}$ for the case when
$n'=0$. Moreover, for this figure we restrict $\Gamma\tau_e$ to $\Gamma\tau_e\in(-\pi/2,\pi)$ for a better overview of possible paths corresponding to these solutions.

\begin{figure}[t!]
   \centering
  \includegraphics[width=10cm]{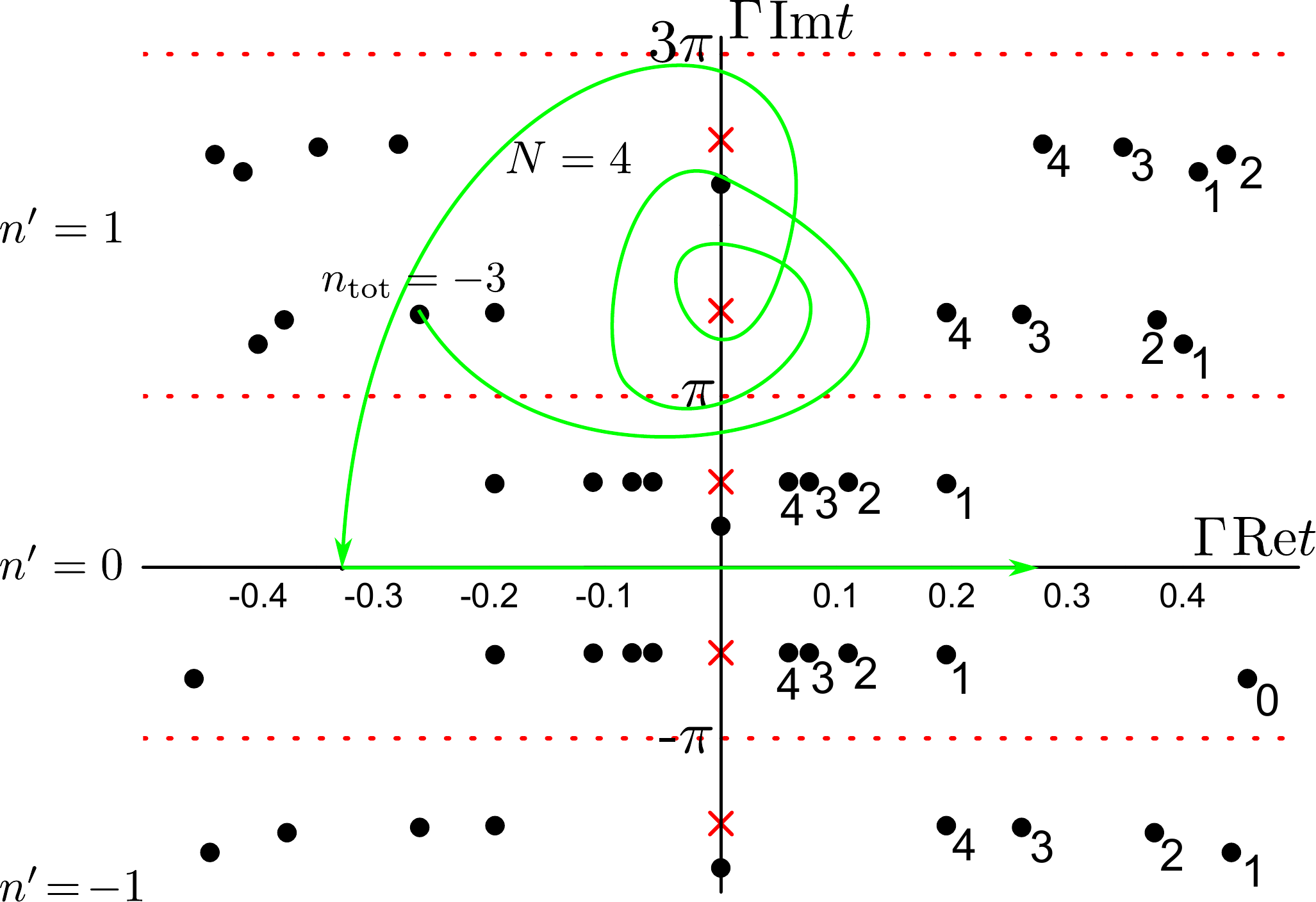}\\
  \caption{Solutions (full black points) of the system \eqref{Eq:first_tau_0_tau_e_full},\eqref{Eq:second_tau_0_tau_e_full} for a wider region in the complex time
  plane than in figure \ref{Fig:paths_encircling}, including $n'\neq 0$. The solutions on the ordinate axis have always $n_\mathrm{tot}=0$.
  The solutions with $\tau_0>0$ are marked by the corresponding values of $n_\mathrm{tot}$ (up to $n_\mathrm{tot}=4$), whereas
  for the solutions with the same absolute values of $\tau_0$ but the opposite sign also the sign of  $n_\mathrm{tot}$ should be flipped respectively.
  The crosses indicate singularity points.
  The shown exemplary path belongs to the solution with $n'=1$, $N=4$ and $n_\mathrm{tot}=-3$.}
  \label{Fig:paths_encircling_ntot}
\end{figure}

In figure \ref{Fig:paths_encircling_ntot} solutions of equations \eqref{Eq:first_tau_0_tau_e_full},\eqref{Eq:second_tau_0_tau_e_full}  are shown
for a larger range of $\Gamma\tau_e$, including also the possibility of $n'\neq 0$. Selecting an arbitrary solution, we indicate a possible path corresponding to this solution so that its characteristic features are clear.

 For paths that may encircle singularities in the complex time plane, in place of equation \eqref{Eq:ImS_solved}  at first we have to use
 \begin{equation}\label{Eq:ImS_solved_full}
 \begin{split}
  \mathrm{Im}\tilde{S}=&\frac{m}{2}
  \int_{0}^{\tau_e}\Big\{\left[\mathrm{Re}\mathcal{V}(\tau_0+i\tau)-\mathrm{Re}\mathcal{V}(t_0)\right]^2-
  \left[\mathrm{Im}\mathcal{V}(\tau_0+i\tau)\right]^2\Big\}\d \tau+\Delta E \tau_e\\
  &+\mathrm{Im}\oint_{\mathring{\delta}}\Big\{F(t)\left[\mathcal{Z}(t,t_0)-\mathrm{Re}\mathcal{V}(t_0)
  (t-t_0)\right]+\frac{m}{2}\left[\mathcal{V}(t)-\mathrm{Re}\mathcal{V}(t_0)\right]^2\Big\}\d t\;.
  \end{split}
\end{equation}
However, at least for the pulse shape given by equation \eqref{Eq:soliton_shape}, we can find that the two terms in the curly brackets of the integral on the second line of equation \eqref{Eq:ImS_solved_full} lead to contributions having opposite signs and eliminating each other. Therefore, equation \eqref{Eq:ImS_solved} may still be used. It can be recast into the following form:
 \begin{equation}\label{Eq:ImS_solved_HSP}
 \begin{split}
  \mathrm{Im}\tilde{S}\frac{2m\Gamma^3}{F_0^2}=&
  \int_{0}^{\Gamma\tau_e}\!\!\!\Big\{\!\! \left[\mathrm{Re}\tanh (\Gamma\tau_0+i\tilde{\tau})\!-\!\mathrm{Re}\tanh (\Gamma\tau_0+i\Gamma\tau_e)\right]^2-\left[\mathrm{Im}\tanh (\Gamma\tau_0+i\tilde{\tau})\right]^2\!\!\Big\}\d \tilde{\tau}\\
 &+\gamma_{\mathrm{HCP}}^2 \Gamma\tau_e.
 \end{split}
\end{equation}


Calculating the probabilities $P$ for various possible $t_0$ in the complex time plane based on equation \eqref{Eq:ImS_solved_HSP}, we
found that all values of $t_0$ with negative imaginary part $\tau_e$ lead to $P>1$. All these solutions have
probability densities rising with the distance inside the barrier. They also have a property that the electron initially moves to the left, away from the barrier, when moving in the complex time plane from the corresponding $t_0$ directly towards the real axis.
Such solution branches are unphysical in the context of the posed tunneling problem and therefore can be ruled out from our present consideration.

\begin{figure}[t!]
   \centering
  \includegraphics[width=12cm]{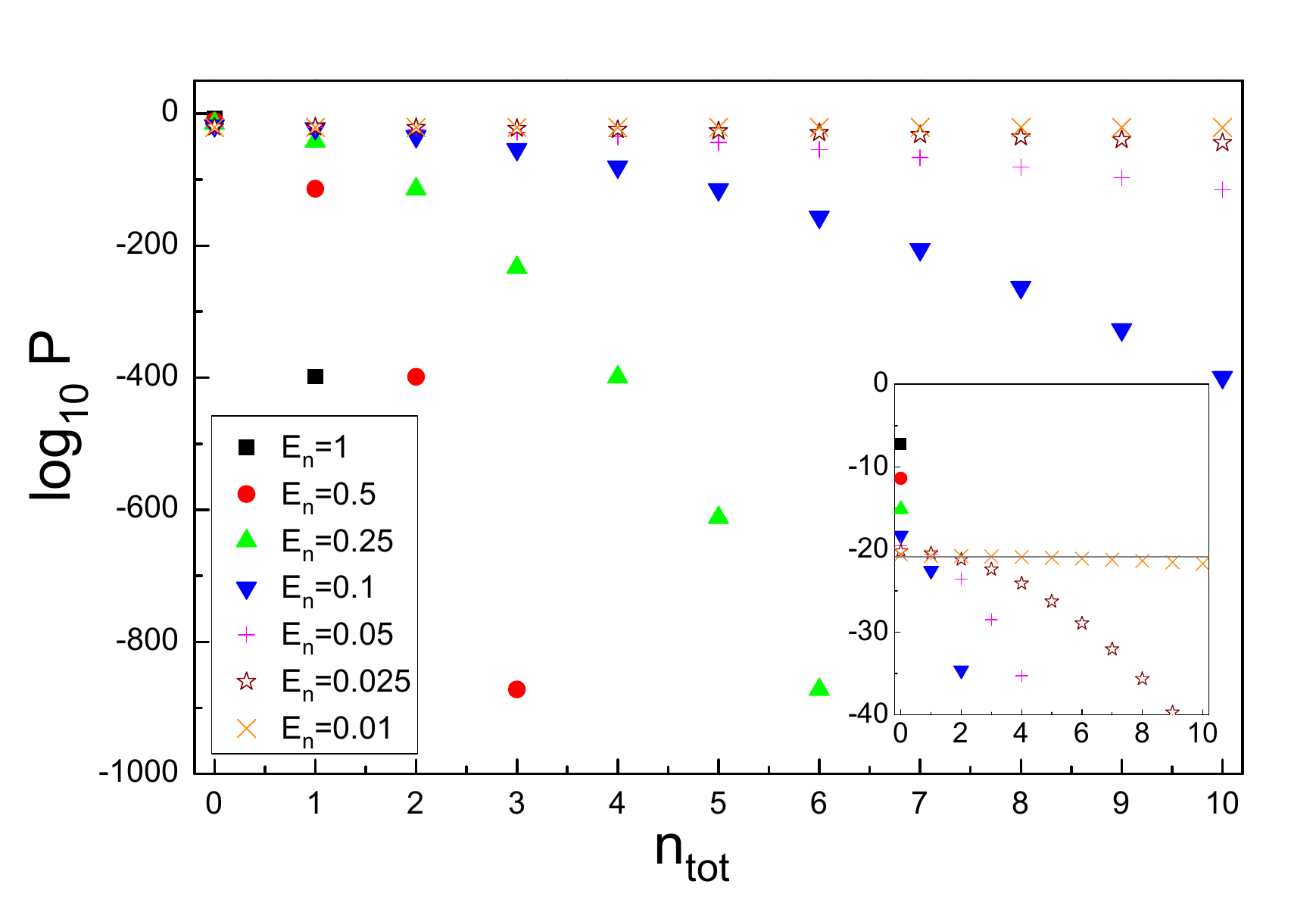}\\
  \caption{Tunneling probabilities for solutions with  $\Gamma\tau_e\in(0,\pi)$ (cf. figure \ref{Fig:paths_encircling}) in dependence on
  $n_\mathrm{not}$
  and the electric field amplitude scaled by a dimensionless factor $E_n$. The probability is determined from equations \eqref{Eq:tun_prob_def} and \eqref{Eq:ImS_solved_HSP} with $\gamma_{_\mathrm{HCP}}=0.94/E_n$ and $F_0^2/(\hbar m \Gamma^3)=34.7E_n^2$, correspondingly.
  The inset shows the magnified region of larger probabilities where $\log_{10} P$ is closer to zero. The black horizontal line in the inset indicate the limit-case value for vanishing electric field after equation \eqref{Eq:validity2}.}
  \label{Fig:paths_convergence}
\end{figure}

For solutions in the upper part of the plane, i.e. with $\tau_e>0$, we found that the maximum value of the tunneling probability corresponds to the solution with $n_\mathrm{tot}=n'=N=0$ having the standard path $\delta_\mathrm{s}$ as a possible integration path in the complex time plane (see blue path in figure \ref{Fig:paths_encircling}).  Typically with increase of the values of $|n_\mathrm{tot}|$ or $n'$ the probability values
drop extremely rapidly with respect to the standard-path solution. In such a situation we can justify neglecting all solutions with non-vanishing  $n_\mathrm{tot}$ or/and $n'$. The situation changes if the respective probabilities do not vary drastically with respect to the standard-path solution. This occurs, e.g., when the amplitude of the driving electric field is decreased, resulting also in the correspondingly increased value of $\gamma_{_\mathrm{HCP}}$. In figure \ref{Fig:paths_convergence} we illustrate how the probabilities corresponding to solutions from the part of the complex time plane shown in figure \ref{Fig:paths_encircling} behave in dependence on $n_\mathrm{tot}$ when the amplitude of the driving field is varied. We can see that for larger electric fields, with $\gamma_{_\mathrm{HCP}}\sim 1$, the change of $n_\mathrm{tot}$ from 0 to 1 already leads to a drop in the probability value as large as several hundred orders of magnitude. However, when the electric field is decreased and the value of the probability of the standard-path ($n_\mathrm{tot}=n'=N=0$)  solution declines towards the value given by equation \eqref{Eq:P_large_gamma} the probabilities of $n_\mathrm{tot}\neq 0$ solutions in contrast rise. For small enough fields, which for our choice of parameters in figure \ref{Fig:paths_convergence} correspond  to $\gamma_{_\mathrm{HCP}}\gtrsim 100$, the dependence of $P(n_\mathrm{tot})$ in the neighbourhood of $n_\mathrm{tot}=0$ saturates to the constant level given by equation \eqref{Eq:P_large_gamma}. In this limit there are many solutions delivering comparable values of the probability. However, within the utilized quasiclassical description there is no straightforward way to combine the corresponding multiple solution branches taking interference effects into account. Since we cannot  use solely the standard-path solution in this case, the validity of the whole approach in its form presented in the current paper breaks down. Where would we then set the validity threshold in terms of the value of $\gamma_{_\mathrm{HCP}}$? Looking at figure \ref{Fig:paths_convergence} we can argue that this threshold is observed when the deviation of the standard-path probability given by equation \eqref{Eq:P_case1} from the value of the limit-case probability given by equation \eqref{Eq:P_large_gamma} is comparable with the latter value. Using for the argument of the exponential in equation \eqref{Eq:P_large_gamma} that for large  $\gamma_{_\mathrm{HCP}}$ we have $\arctan \gamma_{_\mathrm{HCP}}= \frac{\pi}{2}-\arctan(1/\gamma_{_\mathrm{HCP}})\approx \frac{\pi}{2}-(1/\gamma_{_\mathrm{HCP}})$, we obtain the validity condition \eqref{Eq:P_large_gamma_validity_threshold}.

Note that this subtle issue had remained insufficiently clarified  in references \cite{Popov2001,Popov2001_JETP,Popov2005,Popov2004}, whereas it was actually addressed in the last published paper of Leonid V. Keldysh \cite{Keldysh_2017} based on the Keldysh-Reiss-Faisal approach, comparable with other quasiclassical descriptions. Moreover the approach of \cite{Keldysh_2017} allowed to obtain appropriate limit case expressions for the tunneling probability. We agree with  Leonid V. Keldysh that \cite{Keldysh_2017}  ``the weak field regime seems to be of more academic interest: for such short pulses, the effect is hardly experimentally observable.'' However, we should remark that for us it has been also important to
resolve any apparent unexplained paradoxes, like finite probability values in the limit of vanishing strengths of the driving field following from equation \eqref{Eq:P_large_gamma}, to ensure the overall consistency of the utilized method.
Finally, one should also mention that whereas the issue connected to multiple solution branches, originating physically from a multiple-reflection behaviour of the electron moving  in the dynamical potential induced by the field, occurs for the pulse shape given by equation \eqref{Eq:soliton_shape} it may be just absent for other pulse shapes \cite{Popov2001_JETP,Keldysh_2017}.

\section{Branches of the main solution for $t_0(z,t)$}\label{App:roots}
\begin{figure}
   \centering
  \includegraphics[width=\textwidth]{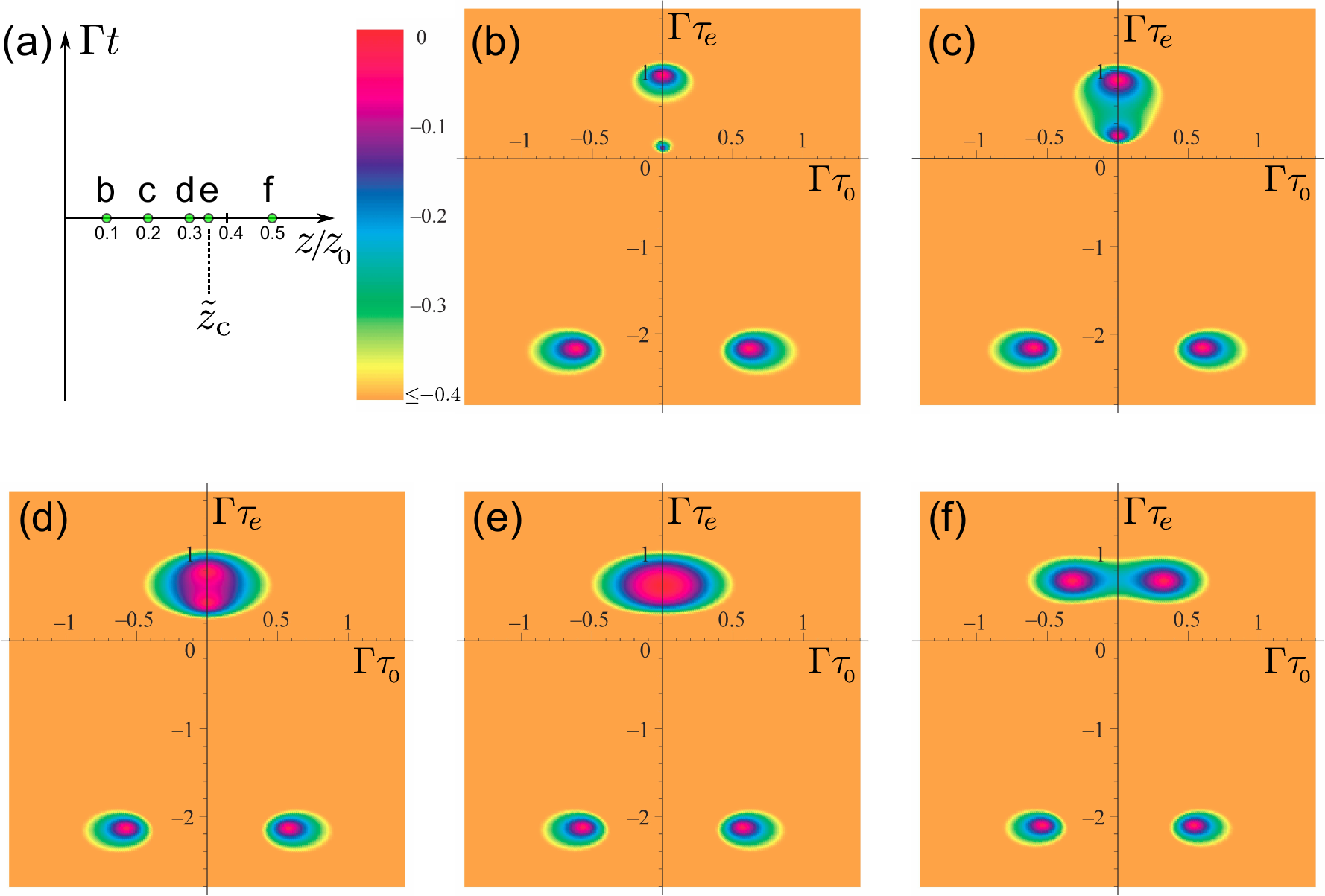}\\
  \caption{Behaviour of  the multiple solutions of equation \eqref{Eq:t0_z_t_sech} is illustrated for the case when the path at $t=0$ along the $z$-axis across the branch point at $z=z_\mathrm{c}$ is chosen. The positions of several points on this path [numbered by letters (b)-(f)] are shown in (a). We used $\gamma_{_\mathrm{HCP}}=0.94$ leading to $\tilde{z}_\mathrm{c}=z_\mathrm{c}/z_0\approx 0.3461$, where $z_0$ is given by equation \eqref{Eq:z0}. For each of the points, the figure numbered by the respective letter shows $-|h(\Gamma t_0)|$ in dependence on $\Gamma\, \mathrm{Re}t_0=\Gamma \tau_0$ and $\Gamma\, \mathrm{Im}t_0=\Gamma \tau_e$, where the function $h(\Gamma t_0)$ is given by equation \eqref{Eq:function_h}.
  Maxima of $-|h(\Gamma t_0)|$ occur exactly when $h(\Gamma t_0)$ vanishes. Their positions (inside of the red spots) determine all possible solutions for $t_0=\tau_0+i \tau_e$ in the shown range for $\Gamma \tau_0$ and $\Gamma \tau_e$: in all cases except of (e)  there are 4 solutions whereas only 3 solutions are present for (e), which corresponds to the branch point.}
  \label{Fig:roots_center}
\end{figure}

Let us illustrate the behaviour of the  branches of the main solution for $t_0=\tau_0+i\tau_e$ determined from equation \eqref{Eq:t0_z_t_sech} in dependence on the final position $z$ and time $t$ considered as the coordinates of the propagating electronic wave packet after the barrier. For the visualization it is convenient to introduce an auxiliary  function
 \begin{equation}\label{Eq:function_h}
  h(\Gamma t_0)=\frac{1}{\Gamma t-\Gamma t_0}\left[\frac{z}{z_0}-\ln\cosh\Gamma t+\ln\cosh \Gamma t_0\right]+\tanh\Gamma t_0 -i\gamma_\mathrm{_{HCP}}
\end{equation}
defined as the difference between the left hand side and the right hand side of equation \eqref{Eq:t0_z_t_sech}. Analyzing the dependence of $-|h(\Gamma t_0)|$ in the complex plane for $t_0$ we can determine the points where it becomes zero, which then means finding the roots of equation \eqref{Eq:t0_z_t_sech}. Moreover, we can study the behaviour  of these points upon varying the final position $z$ and time $t$. In figure \ref{Fig:roots_center} we fix the final time at $t=0$, corresponding to the peak value of the driving electric field, and change the position $z$. Several selected points situated along the corresponding path in the plane $(z,t)$ and labeled as (b)-(f) are shown in figure \ref{Fig:roots_center}(a). The behaviour of $-|h(\Gamma t_0)|$ in the plane $\tau_0,\tau_e$ at each of these points is then illustrated in the corresponding figures  \ref{Fig:roots_center}(b)-(f).

\begin{figure}
   \centering
  \includegraphics[width=\textwidth]{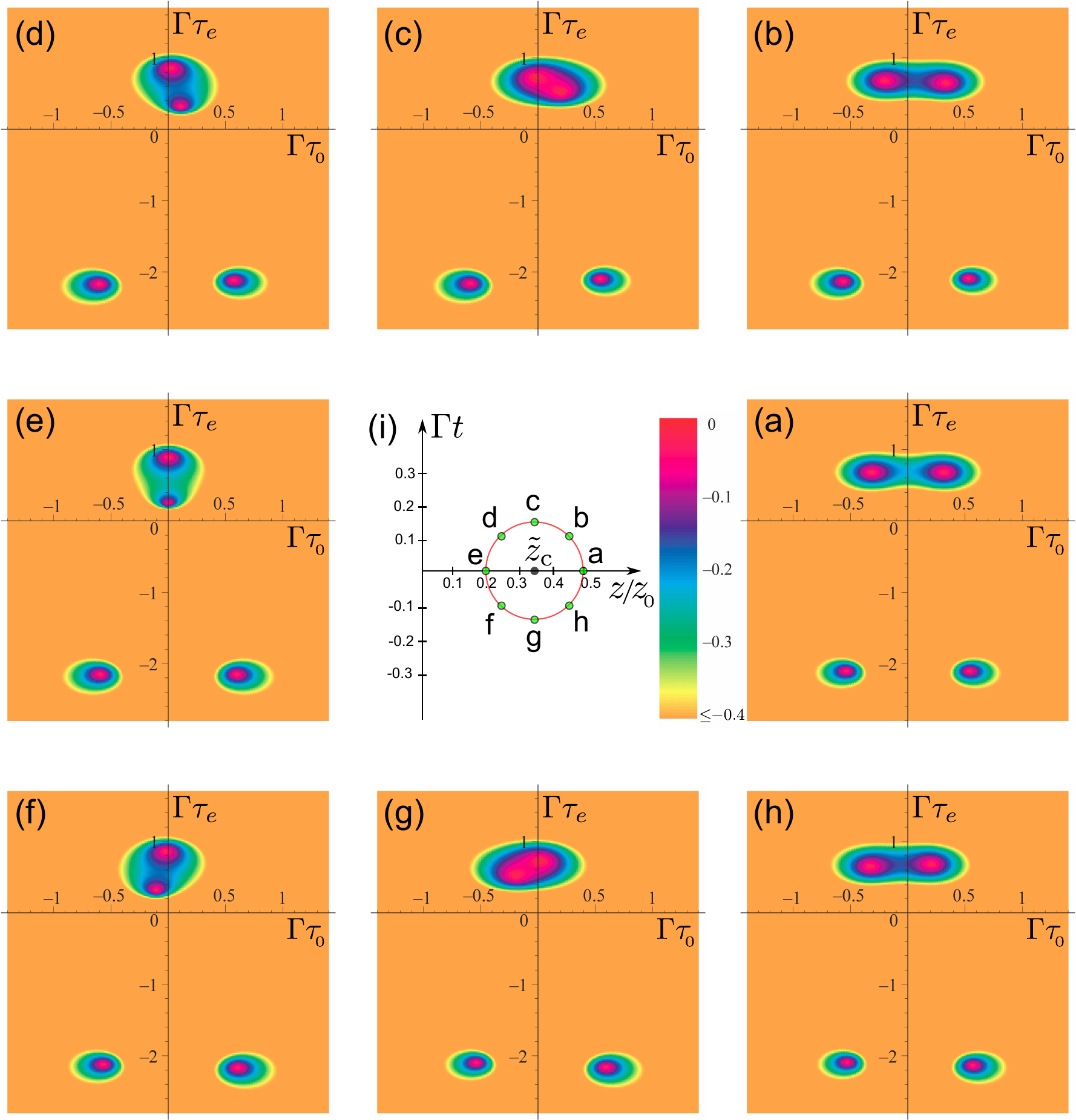}\\
  \caption{Behaviour of  the multiple solutions of equation \eqref{Eq:t0_z_t_sech} illustrated similar to figure \ref{Fig:roots_center} but now selecting a path around the branch point located at $(z=z_\mathrm{c},t=0)$.}
  \label{Fig:roots_circle}
\end{figure}

As we noticed in \ref{App:paths}, the  solutions in the lower part of the complex time plane are unphysical and may be ignored. In figures \ref{Fig:roots_center}(b)-(f) we observe that there are two main solutions, which are imaginary for small $z$. With increase of $z$ these solutions move towards each other along the real time axis until they merge at $z=z_\mathrm{c}$. Then with further increase of $z$ they again split and   start to move parallel to the real time axis away from each other [see also figure \ref{Fig:bifurcation_plot}(a)]. Having the same imaginary part of $t_0$, these two solutions then lead to the same value of the probability at $t=0$ and large $z$ [cf. figure \ref{Fig:bifurcation_plot}(a)]. In fact, as we can clearly see from figure \ref{Fig:roots_circle}, these are two branches of the same multivalue solution. The solution branches go over into each other [see figures \ref{Fig:roots_circle}(a)-(h)] by performing a full turn along a circular path around the branch point at $(z=z_\mathrm{c},t=0)$  [see figure  \ref{Fig:roots_circle}(i)] .

\section*{References} \addcontentsline{toc}{section}{References}
\bibliographystyle{iopart-num}
\providecommand{\newblock}{}

\end{document}